\documentclass[aps,prb,superscriptaddress,amsfonts,amsmath,amssymb,twocolumn,floatfix]{revtex4}
\usepackage{bm}
\usepackage{graphicx}
\usepackage{amsmath}
\usepackage{amstext}
\usepackage{amssymb}
\usepackage{amsfonts}
\usepackage{amsbsy}
\usepackage{verbatim}

\def\bx{\boldsymbol{x}}
\def\by{\boldsymbol{y}}
\def\ba{\boldsymbol{a}}

\def\br{\boldsymbol{r}}
\def\bR{\boldsymbol{R}}
\def\vR{\vec{R}}
\def\bk{\boldsymbol{k}}

\def\bq{\boldsymbol{q}}
\def\bQ{\boldsymbol{Q}}
\def\bS{\boldsymbol{S}}

\def\bN{\boldsymbol{N}}
\def\bn{\boldsymbol{n}}

\def\bC{\boldsymbol{C}}

\def\bz{\boldsymbol{z}}

\def\rSU{{\rm SU}}

\def\bsigma{\boldsymbol{\sigma}}

\def\rSU2{{\rm SU}(2)}

\begin{document}

\title{Properties of an algebraic spin liquid on the kagome lattice}

\author{Michael Hermele}
\affiliation{Department of Physics, University of Colorado, Boulder, Colorado 80309, USA}
\author{Ying Ran}
\affiliation{Department of Physics, University of California, Berkeley, California 94720, USA}
\author{Patrick A. Lee}
\affiliation{Department of Physics, Massachusetts Institute of Technology, Cambridge, Massachusetts 02139, USA}
\author{Xiao-Gang Wen}
\affiliation{Department of Physics, Massachusetts Institute of Technology, Cambridge, Massachusetts 02139, USA}

\date{\today}
\begin{abstract}
In recent work, we argued that a particular algebraic spin liquid (ASL) may be the ground
state of the $S = 1/2$ kagome lattice Heisenberg antiferromagnet. Furthermore, this state, which
lacks a spin gap, is appealing in light of recent experiments on herbertsmithite (ZnCu$_3$(OH)$_6$Cl$_2$).
Here, we study the properties of this ASL in more detail, using both the low-energy effective field
theory and Gutzwiller-projected wavefunctions of fermionic spinons. We identify the competing
orders of the ASL, which are observables having slowly-decaying power law correlations -- among
them we find a set of magnetic orders lying at the M-points of the Brillouin zone, the familiar
$\bq = 0$ magnetic ordered state, the ``Hastings'' valence-bond solid (VBS) state, and a pattern of
vector spin chirality ordering corresponding to one of the Dzyaloshinskii-Moriya (DM) interaction
terms present in herbertsmithite. Identification of some of these orders requires an understanding
of the quantum numbers of magnetic monopole operators in the ASL. We discuss the detection
of the magnetic and VBS competing orders in experiments.  While we primarily focus on a clean
system without DM interaction, we consider the effects of small DM interaction and argue that,
surprisingly, it leads to spontaneously broken time reversal symmetry (for DM interaction that
preserves XY spin rotation symmetry, there is also XY magnetic order). Our analysis
of the projected wavefunction provides an estimate of the ``Fermi velocity'' $v_F$ that characterizes
all low-energy excitations of the ASL -- this allows us to estimate the specific heat, which compares
favorably with experiments. We also study the spin and bond correlations of the projected
wavefunction and compare these results with those of the effective field theory. While the spin
correlations in the effective field theory and wavefunction seem to match rather well (although not
completely), the bond correlations are more puzzling. We conclude with a discussion of experiments
in herbertsmithite and make several predictions.
\end{abstract}
\maketitle

\section{Introduction}
\label{sec:intro}

Recent experiments on herbertsmithite (ZnCu$_3$(OH)$_6$Cl$_2$) have reinvigorated  the longstanding interest in the ground state of the $S = 1/2$ Heisenberg antiferromagnet on the kagome lattice.\cite{helton07,mendels07,ofer06, imai08, shlee07, devries07, bert07, olariu08}  This material contains kagome layers of antiferromagnetically coupled Cu$^{2+}$ $S = 1/2$ moments, with an exchange energy $J \approx 200 {\rm K}$.  The coupling between adjacent kagome layers is expected to be very small.  Remarkably, no sign of ordering -- magnetic or otherwise -- has been observed down to the lowest temperatures measured ($50 {\rm mK}$ for some probes).  Frozen magnetic moments and spin glass behavior are also not observed.  Herbertsmithite is thus a candidate for the experimental realization of a quantum spin liquid in two dimensions.\cite{anderson73, anderson87}

Various experimental probes point to a vanishing spin gap.  Spin liquids with this property are variously (and equivalently) referred to as gapless spin liquids, critical spin liquids, or long-range resonating valence bond (RVB) states.  So far, most works studying nonmagnetic ground states of the kagome antiferromagnet have focused on gapped spin liquid states,\cite{marston91,sachdev92} or valence-bond solid (VBS) states that break lattice symmetries.\cite{marston91, hastings00,nikolic03}  A variety of microscopic calculations have provided interesting information, but are unable to establish the nature of the ground state.\cite{leung93, elstner94, zeng95,  lecheminant97, waldtmann98, sindzingre00, misguich07, singh07, singh08, yang08}  (See also Ref.~\onlinecite{misguich05} for a review.)

The experimental work on herbertsmithite led us, in recent work, to investigate the possibility of a gapless spin liquid ground state in the kagome lattice Heisenberg model.\cite{ran07}  By considering a class of Gutzwiller-projected fermion wavefunctions, we concluded that the variational ground state of the kagome model is a particular kind of gapless spin liquid known as an algebraic spin liquid (ASL).\cite{rantner01, rantner02}  Some of us studied the effect of a Zeeman magnetic field, and argued that it leads to spontaneous breaking of parity and XY spin rotation.\cite{ran07b}  Gregor and Motrunich considered the effect of non-magnetic impurities in the ASL, finding results consistent with the NMR experiments on herbertsmithite.\cite{gregor08}
In this paper, we shall work out the properties of this realization of the ASL in more detail; this leads to a number of predictions that may be relevant for herbertsmithite.  

We note that Ma and Marston have considered a different gapless spin liquid (Fermi surface state) using Gutzwiller projected wavefunctions, and have argued it can be stabilized by addition of further-neighbor ferromagnetic exchange.\cite{ma08}   Also, a different route to a gapless spin liquid on the kagome lattice has recently been discussed by Ryu \emph{et. al.}\cite{ryu07}  

The effective field theory describing the ASL consists of $N_f = 4$ flavors of massless, two-component Dirac fermions coupled to a ${\rm U}(1)$ gauge field.  This description is complementary to the projected wavefunction approach -- the former correctly captures universal long-distance properties, while the latter can provide information about energetics and other more microscopic properties.  Significant progress has been made recently in understanding algebraic spin liquids using effective field theory,\cite{rantner01, rantner02, borokhov02, vafek02, hermele04, hermele05, alicea05a, alicea05b, alicea06, ryu07} and it has been found that such states exhibit striking observable properties.  The ASL is a quantum critical \emph{phase} -- like a quantum critical point, it supports gapless excitations and nontrivial critical exponents, but can exist as a stable zero-temperature phase and can be accessed with no fine tuning of parameters.  The ASL is Lorentz invariant, and all excitations propagate with the same ``Fermi velocity'' $v_F$.  The symmetry of the ASL is much larger than that of the microscopic spin model, and this emergent symmetry unifies together a variety of superficially unrelated observables.  Some of these observables have slowly decaying power law correlations in space and time; these are referred to as ``competing orders."

In this paper, we show how the observable properties of the ASL are manifested in the kagome lattice model.  In particular, we identify the competing orders most likely to have slowly decaying correlations -- these include magnetic orders and valence-bond solid (VBS) states, as well as patterns of order involving vector and scalar spin chiralities.  We discuss the detection of the magnetic and VBS orders in experiments.  We also present further results in the projected wavefunction approach -- in particular, we give an estimate of the velocity $v_F$, and study the spin-spin and bond-bond correlations of the wavefunction.  Taken together, these results inform a variety of predictions that may be relevant for herbertsmithite.  Furthermore, the estimate of $v_F$ allows us to estimate the low-temperature specific heat and magnetic susceptibility, and we find reasonable agreement with experiments.

It bears mentioning that herbertsmithite is almost certainly not described by a Heisenberg model alone.  There is now significant evidence that impurities play an important role, especially at low temperature.\cite{imai08,devries07,bert07,olariu08}  It has been suggested that antisite defects, where Zn and Cu trade places, constitute the dominant disorder -- this leads to both magnetic impurities (Cu occupying Zn sites) and site dilution in the kagome layers (Zn occupying Cu sites).\cite{devries07}  Estimates of the concentration of magnetic impurities per kagome lattice site are in the range of $5\%$ to $10\%$.  It has also been suggested that Dzyaloshinskii-Moriya (DM) interaction is an important perturbation to the Heisenberg model.\cite{rigol07a, rigol07b}  While DM interaction is certainly present, its magnitude is uncertain.

Here, we have not attempted to address the effects of impurities and DM interaction in detail.  Instead, our approach is to first understand the spin liquid physics in a clean model with only Heisenberg exchange.  One can then include impurities and DM as perturbations to this idealized model.  While we do discuss the effects of small DM interaction, the effects of impurities, and of strong DM interaction, are left for future work.

We now give an outline of the paper, which also serves as a more detailed overview of our main results.  Readers primarily interested in predictions for experiment can skip Secs.~\ref{sec:description} and~\ref{sec:many-competing-orders}.  We begin in Sec.~\ref{sec:effective-theory} by discussing how the effective field theory is obtained starting from the Heisenberg model.  The field theory can be analyzed in a large-$N_f$ expansion, and our understanding of the physical case ($N_f = 4$) derives from the analysis in this limit.  In Sec.~\ref{sec:symmetries}, we show how the microscopic symmetries are realized as transformations on the fields of the continuum theory.

Section~\ref{sec:many-competing-orders} describes our main results from the effective field theory approach.  Section~\ref{sec:co-gendisc} reviews the physics of competing orders in the ASL.  There are two sets of field theory operators whose correlations are likely to decay slowly for $N_f = 4$.  These are 15 fermion bilinears, and a set of 12 magnetic monopole operators.  In Sec.~\ref{sec:fermion-bilinears} we relate each of the fermion bilinears to an ordering pattern in the kagome spin model.  The ordering patterns arising are the valence-bond solid (VBS) state considered before by Hastings,\cite{hastings00}  a set of magnetic orders with crystal momenta lying at the M-points in the Brillouin zone, and a pattern of vector spin chirality ordering, which also corresponds to one of the DM terms allowed by symmetry in herbertsmithite.  In Sec.~\ref{sec:monopoles}, we relate the monopole operators to ordering patterns in the spin model.  This is less straightforward than the analysis for the fermion bilinears, as there is an ambiguity in determining the symmetry transformations of the monopoles.  This ambiguity is discussed, and partially resolved using algebraic relations among generators of the space group.\cite{alicea05a, alicea05b, alicea06, ryu07}  Because the ambiguity in monopole transformations can only be partially resolved, we are left with three free parameters describing the possible transformation laws.  We make a conjecture on the value of two of these parameters; based on this, we find among the monopoles a pattern of magnetic order corresponding to the familiar ``$\bq = 0$'' state of the kagome lattice.  The conjecture is supported by the spin correlations of the projected wavefunction (Sec.~\ref{sec:spin-correlation}).  Depending on the remaining parameter, there are monopoles corresponding to either
a pattern of spin-chirality ordering that breaks time reversal symmetry, or a VBS ordering pattern.

In Sec.~\ref{sec:detection}, we discuss how to detect some of the competing orders in experiment.  In particular, the M-point and $\bq = 0$ magnetic orders can be detected via NMR relaxation rate, and also by inelastic neutron scattering -- both these quantities obey universal scaling forms determined by the critical properties of the ASL.  The VBS competing order can be detected via its coupling to an appropriate optical phonon.  The lineshape of this phonon can be related to the VBS susceptibility of the ASL, which again obeys a scaling form as a function of frequency and temperature.  

In Sec.~\ref{sec:dm-interaction}, we consider the effect of a small DM interaction.  Considering first DM interactions that preserve XY spin rotation symmetry, we argue that, surprisingly, DM interaction induces XY magnetic ordering, which is likely to be in the $\bq = 0$ pattern.  More generically, DM interaction completely breaks spin rotation symmetry, and in this case the same argument shows that it leads to spontaneous breaking of time reversal.

In Sec.~\ref{sec:analysis-pw} we turn to a further analysis of the projected wavefunction.  By construction of variational excited states, in Sec.~\ref{sec:fermi-velocity} we estimate the velocity $v_F$.  Later on in Sec.~\ref{sec:discussion}, this allows us to calculate the specific heat and magnetic susceptibility in mean-field theory; these quantities compare favorably with the experimental data.  In Secs.~\ref{sec:spin-correlation} and~\ref{sec:bond-correlation} we investigate the spin-spin and bond-bond correlators of the projected wavefunction.  The spin-spin correlator is dominated by correlations in the 
pattern of the $\bq = 0$ state -- in real space these correlations fall off roughly as $1/r^4$.  Although the correlations of the M-point order are expected to decay more slowly than this, they are not seen; this may indicate that these correlations have a small coefficient and thus only become important at very long distances.  The bond-bond correlation function exhibits power law decay, but the observed correlations 
are apparently dominated by Fourier components near $\bq = 0$ and thus do not correspond to the Hastings VBS state (or to the VBS state that may arise from the monopoles, which also has crystal momenta at the M-points).  We speculate on the possible meaning of this in Sec.~\ref{sec:bond-correlation}.

We conclude in Sec.~\ref{sec:discussion} with a discussion of experiments on herbertsmithite, and open theoretical issues.  Various technical details are contained in the appendices.

\section{Description of algebraic spin liquid}
\label{sec:description}

\subsection{Effective theory}
\label{sec:effective-theory}

\begin{figure}
\includegraphics[width=3in]{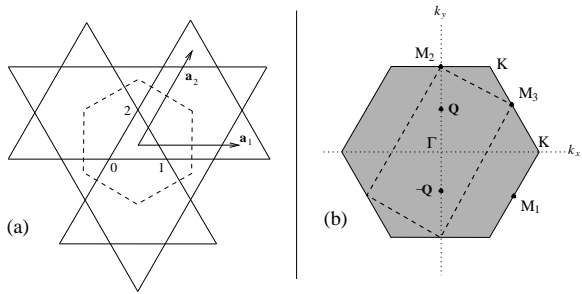}
\caption{(a) Unit cell of the kagome lattice, showing the lattice vectors $\ba_1$ and $\ba_2$, and the numbering of the three sites within the unit cell.  (b)  Brillouin zone.  The M, K and $\Gamma$ points are labeled.  The M points will play an important role in our discussion, and have thus been numbered as shown to distinguish among them.  The reduced Brillouin zone of the enlarged real-space unit cell (Fig.~\ref{fig:big-unitcell}) is denoted by the dashed rectangle.  The Dirac nodes lie at $\pm \bQ$, where $\bQ = (\pi / \sqrt{3})\by$.}
\label{fig:unitcell-and-bz}
\end{figure}

We are interested in the $S = 1/2$ Heisenberg antiferromagnet on the kagome lattice, with Hamiltonian
\begin{equation}
\label{eqn:heisenberg-hamiltonian}
{\cal H} = J \sum_{\langle \br \br' \rangle} \bS_{\br} \cdot \bS_{\br'} \text{,}
\end{equation}
where the sum  is over nearest-neighbor pairs of sites.  The kagome lattice has a three-site unit cell; we label the sites by pairs $(\bR, i)$, where $\bR = n_1 \ba_1 + n_2 \ba_2$ is the lattice vector labeling the unit cell, and $i = 0, 1, 2$ labels the three sites (Fig.~\ref{fig:unitcell-and-bz}a).  We choose $\ba_1= \bx$ and $\ba_2 = (1/2)\bx + (\sqrt{3}/2) \by$, so the distance between nearest neighbor sites is $1/2$.  The reciprocal lattice primitive vectors can be chosen as $\boldsymbol{b}_1 = 2\pi [ \bx - (1/\sqrt{3}) \by]$ and 
$\boldsymbol{b}_2 = ( 4 \pi / \sqrt{3} ) \by$, and the Brillouin zone is as shown in Fig.~\ref{fig:unitcell-and-bz}b.

Although our main focus is on the pure Heisenberg model, we shall also consider Dzyaloshinskii-Moriya (DM) interactions, of the type allowed by the crystal symmetries in ZnCu$_3$(OH)$_6$Cl$_2$.  We refer the reader to Ref.~\onlinecite{rigol07b} for the definition of the two allowed DM terms.  In one of these, the $D_z$-term, the DM vector points along the $z$-axis.  The $D_p$-term, on the other hand, has DM vectors lying in the $xy$-plane.

We begin by representing the single-site Hilbert space in terms of $S = 1/2$ fermionic spinons:
\begin{equation}
| \alpha \rangle  = f^\dagger_{\alpha} | 0 \rangle \text{,}
\end{equation}
where $\alpha = \uparrow, \downarrow$.  We choose fermions (as opposed to bosons) because they allow for the description of stable, gapless spin liquid phases.  This representation involves an enlargement of the Hilbert space, and must be accompanied by the local constraint $f^\dagger_{\br \alpha} f^{\vphantom\dagger}_{\br \alpha} = 1$ to eliminate unphysical empty and doubly occupied sites.  

These variables allow one to construct low-energy effective theories for spin liquid phases, as well as corresponding trial ground state wavefunctions.  The starting point for these constructions is a decoupling of the quartic exchange interaction using an auxiliary field residing on the bonds of the lattice.  Neglecting the fluctuations of the auxiliary field, one arrives at a quadratic mean-field spinon Hamiltonian.  In order to obtain a correct description of any spin liquid ground state, one needs to go beyond the mean-field description and consider the fluctuations of the auxiliary field, which play an important role.  One way to do this is to solve the mean-field Hamiltonian, and then perform a Gutzwiller projection onto the subspace with one fermion per site; this results in a legitimate trial wavefunction for the spin model.  Alternatively, one can recognize that the fluctuations about the mean-field saddle point take the form of a gauge field coupled to the spinons.  One can then write down an effective gauge theory Hamiltonian, which will correctly capture the universal features of a given spin liquid phase.

In a recent paper, we studied the kagome antiferromagnet using Gutzwiller projected wavefunctions.\cite{ran07}  The main result of this study was that a particular spin liquid state, first discussed by Hastings,\cite{hastings00} has a very low energy, even without any tuning of variational parameters.  This state has the mean-field Hamiltonian
\begin{equation}
\label{eqn:mft-hamiltonian}
{\cal H}_{{\rm MFT}} = - t \sum_{\langle \br \br' \rangle} s_{\br \br'} \big( f^\dagger_{\br \alpha} f^{\vphantom\dagger}_{\br' \alpha} + \text{H.c.} \big) \text{.}
\end{equation}
The hopping parameter $t$ is sometimes written $t = \chi J$, so that $\chi$ gives the mean-field hopping in units of $J$.  Also, $s_{\br \br'} = \pm 1$ encodes the background magnetic flux of the gauge field coupled to the spinons; it is chosen so that $\pi$-flux pierces the kagome hexagons, and zero flux pierces the triangular plaquettes.  The total number of spinons is chosen so that $\langle f^\dagger_{\br \alpha} f^{\vphantom\dagger}_{\br \alpha} \rangle  =1$.  

\begin{figure}
\includegraphics[width=2.8in]{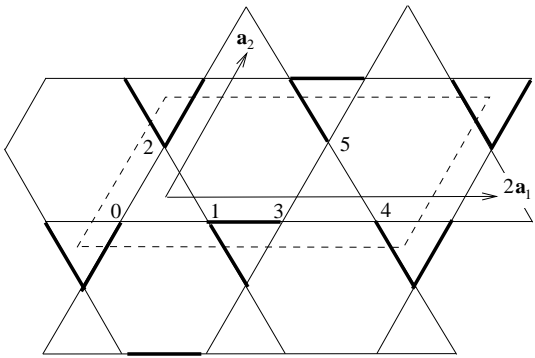}
\caption{The doubled unit cell used to diagonalize the mean-field spinon Hamiltonian Eq.~(\ref{eqn:mft-hamiltonian}).  The lattice vectors are $2 \ba_1$ and $\ba_2$, as shown.  It should be stressed that the unit cell doubling is a gauge artifact, and does not represent a breaking of translation symmetry.  The thick bonds are those where $s_{\br \br} = -1$, and the sites are numbered as shown.}
\label{fig:big-unitcell}
\end{figure}

Instead of a Fermi surface, ${\cal H}_{{\rm MFT}}$ has gapless Dirac points at the Fermi energy, near which the fermions obey a massless Dirac dispersion with velocity $v_F$.  One can diagonalize the Hamiltonian using the 6-site unit cell of Fig.~\ref{fig:big-unitcell}, with the signs of $s_{\br \br'}$ chosen as shown; the Dirac nodes lie in the reduced Brillouin zone at positions $\pm \bQ$, where $\bQ = \pi \by / \sqrt{3}$ (Fig.~\ref{fig:unitcell-and-bz}).  One can describe the low-energy excitations near the nodes in terms of the Lagrange density for Dirac fermions in $2+1$-dimensional spacetime:
\begin{equation}
{\cal L}_{{\rm MFT}} = \bar{\psi}_{\alpha a} \big[ -i \gamma_{\mu} \partial_{\mu} \big] \psi_{\alpha a} \text{.}
\end{equation}
Here, $\psi_{\alpha a}$ is a two-component fermion field, with $\alpha = \uparrow,\downarrow$ labeling the spin, and $a = +, -$ labeling the two nodes at $\bQ$ and $- \bQ$, respectively.  The two components of $\psi_{\alpha a}$ correspond to the two branches of the Dirac dispersion.  Moreover, the index $\mu = 0,1,2$, and $\gamma_{\mu} = (\tau^3, \tau^2, -\tau^1)$, where the $\tau^i$ are $2 \times 2$ Pauli matrices.  Finally, $\bar{\psi}_{\alpha a} \equiv i \psi^\dagger \tau^3$.  More details on the band structure of ${\cal H}_{{\rm MFT}}$, as well as its continuum limit, and the detailed relationship between $\psi_{\alpha a}$ and the lattice spinon fields, are given in Appendix~\ref{app:cont-limit}.

The important fluctuations about the mean-field saddle point specified by ${\cal H}_{{\rm MFT}}$ are encapsulated in the coupling of the spinons to a compact ${\rm U}(1)$ gauge field, and the spin liquid is thus an algebraic spin liquid.  Because of this structure, this state has been referred to as the ${\rm U}(1)$-Dirac state;\cite{ran07,ran07b} here, however, to emphasize the connections with earlier work we shall refer to it as an algebraic spin liquid. The coupling to the gauge field is encoded in the following lattice gauge theory Hamiltonian:
\begin{eqnarray}
\label{eqn:effective-lattice-hamiltonian}
{\cal H}_{{\rm eff}} &=& h \sum_{\langle \br \br' \rangle} e^2_{\br \br'} - K \sum_p \cos \big( [\nabla \times a]_p \big) \nonumber \\
&-& t \sum_{\langle \br \br' \rangle} s_{\br \br'} \big( f^\dagger_{\br \alpha} e^{i a_{\br \br'}} f^{\vphantom\dagger}_{\br' \alpha} + \text{H.c.} \big) \text{.}
\end{eqnarray}
Here $e_{\br \br'}$ and $a_{\br \br'}$ are the lattice electric field and vector potential, respectively.  They reside on nearest-neighbor bonds of the lattice and satisfy the canonical commutation relation $[a, e] = i$.  The electric field has integer-valued eigenvalues, and the vector potential's eigenvalues are $2\pi$-periodic and can be taken in the interval $[-\pi, \pi)$.  The notation $\sum_p$ denotes a sum over all triangular and hexagonal kagome lattice plaquettes, and $[\nabla \times a]_p$ denotes the discrete (oriented) line integral of $a_{\br \br'}$ around the corresponding plaquette, \emph{i.e.} the magnetic flux of the gauge field.  In general we should allow the energy $K$ to differ on triangular and hexagonal plaquettes, but this will not be important for our discussion here.  This Hamiltonian must be supplemented by the Gauss' law constraint
\begin{equation}
(\nabla \cdot e)_{\br}  = f^\dagger_{\br \alpha} f^{\vphantom\dagger}_{\br \alpha} - 1 \text{,}
\end{equation}
where $(\nabla \cdot e)_{\br} \equiv \sum_{\br' \sim \br} e_{\br \br'}$  is the lattice divergence of the electric field (the sum is over nearest neighbors $\br'$ of $\br$).  

In the limit $h \gg t, K$, one recovers a spin model Hamiltonian.  However, this effective description is most useful in the limit of large $K$, where fluctuations of the magnetic field are suppressed and the spinons become good variables to describe the physics (at least for short length scales).  The short-distance physics of the Heisenberg model is presumably \emph{not} similar to the short-distance physics of ${\cal H}_{{\rm eff}}$ in the large-$K$ limit.  Instead, the idea is that the two Hamiltonians may have the same long-distance physics, \emph{i.e.} they are in the same phase.  

In the large-$K$ limit one obtains the algebraic spin liquid, which is described by the Lagrangian density
\begin{eqnarray}
\label{eqn:effective-lagrangian}
{\cal L}_{{\rm eff}}  &=& \bar{\psi}_{\alpha a} \big[ -i \gamma_{\mu} ( \partial_{\mu} 
+ i a_{\mu} ) \big] \psi_{\alpha a} \nonumber \\
&+& \frac{1}{2 e^2} \sum_{\mu} (\epsilon_{\mu \nu \lambda} \partial_{\nu} a_{\lambda})^2 + \cdots   \text{.}
\end{eqnarray}
This is the so-called QED3 Lagrangian, which includes the minimal coupling of the gauge field to the spinons, as well as an explicit Maxwell term for the gauge field.  In general we may (and must) add other perturbations as allowed by microscopic symmetries -- such terms are represented by the ellipsis.  While QED3 is a strongly coupled problem, in the sense that the interaction between spinons and gauge field is strongly relevant in the RG sense, it can be understood in a large $N_f$ limit, where the number of two-component fermions fields is increased from 4 to $N_f$ (\emph{e.g.} $\alpha = 1, \dots, N_f/2$).  The theory can be solved for $N_f \to \infty$, and one can calculate perturbatively in powers of $1/N_f$.  This large-$N_f$ expansion underpins the understanding of the algebraic spin liquid fixed point, and has been discussed in great detail elsewhere.\cite{appelquist86, borokhov02, franz02, rantner02, hermele04, hermele05}

It shall be convenient to organize the four two-component fermions into the eight-component object
\begin{equation}
\Psi = \left( \begin{array}{c}
\psi_{\uparrow, +} \\
\psi_{\uparrow, -} \\
\psi_{\downarrow, +} \\
\psi_{\downarrow, -}
\end{array} \right) \text{.}
\end{equation}
The $\tau^i$ Pauli matrices act in the two-component space of each Dirac fermion, so that
\begin{equation}
\tau^i \Psi = \left( \begin{array}{c}
\tau^i \psi_{\uparrow, +} \\
\tau^i \psi_{\uparrow, -} \\
\tau^i \psi_{\downarrow, +} \\
\tau^i \psi_{\downarrow, -}
\end{array} \right) \text{.}
\end{equation}
We also define $\sigma^i$ Pauli matrices acting in the spin space, and $\mu^i$ Pauli matrices acting in the ``nodal'' space connecting the two nodes at $\pm \bQ$.  For example, we have
\begin{equation}
\sigma^3 \Psi = \left( \begin{array}{c}
\psi_{\uparrow, +} \\
\psi_{\uparrow, -} \\
- \psi_{\downarrow, +} \\
- \psi_{\downarrow, -}
\end{array} \right) \text{,}
\end{equation}
and
\begin{equation}
\mu^3 \Psi = \left( \begin{array}{c}
\psi_{\uparrow, +} \\
- \psi_{\uparrow, -} \\
\psi_{\downarrow, +} \\
- \psi_{\downarrow, -}
\end{array} \right) \text{.}
\end{equation}
Different types of Pauli matrices commute with one another:
\begin{equation}
\left[ \sigma^i, \mu^j \right] = \left[ \mu^i, \tau^j \right] = \left[ \tau^i, \sigma^j \right] = 0 \text{.}
\end{equation}

\subsection{Symmetries}
\label{sec:symmetries}

\begin{figure}
\includegraphics[width=2.5in]{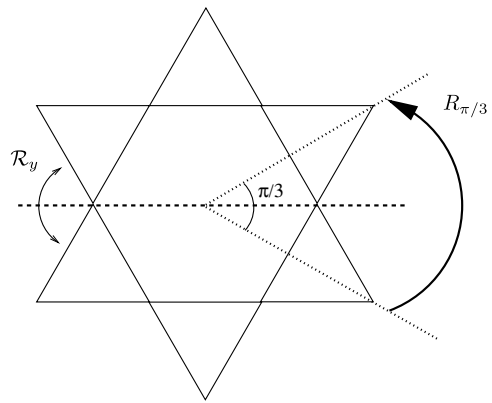}
\caption{Depiction of the generators of the point group symmetry of the kagome lattice, ${\cal R}_y$ and $R_{\pi/3}$.}
\label{fig:point-symmetries}
\end{figure}

For our analysis later on, we need to work out how the symmetries of the microscopic Hamiltonian Eq.~(\ref{eqn:heisenberg-hamiltonian}) are realized in the continuum theory.  The procedure for obtaining these results is described in Appendix~\ref{app:symmetries}; here, we shall simply enumerate the symmetries and quote the results.  The space group of the kagome lattice is generated by translations by the Bravais lattice vector $\bR = n_1 \ba_1 + n_2 \ba_2$ ($T_{\bR}$), rotations by $\pi/3$ about the center of a hexagonal plaquette ($R_{\pi/3}$), and a reflection ${\cal R}_y$ mapping $y \to -y$.  These point group symmetries are depicted in Fig.~\ref{fig:point-symmetries}.  For definiteness in defining the rotations and reflections, we take the origin of coordinates at the center of a hexagon.  The remaining symmetries are spin rotations and time reversal.

For each symmetry operation, its action on the lattice spinon fields $f_{\br \alpha}$ is determined by two requirements.  First, it must reproduce the correct action of the symmetry on the spin operator $\bS_{\br}$ -- this ensures that the action of the symmetry on all physical operators is correctly realized.  This requirement does not completely specify the action of the symmetry on $f_{\br \alpha}$, because $\bS_{\br}$ is invariant under local ${\rm SU}(2)$ gauge transformations of the spinons -- the symmetry may therefore be supplemented by an \emph{arbitrary} ${\rm SU}(2)$ gauge transformation.  However, we should also impose a second requirement, which is that the effective Hamiltonian Eq.~(\ref{eqn:effective-lattice-hamiltonian}) be invariant.  In the present case, this requirement will fix the symmetry operation completely up to multiplication by a global ${\rm U}(1)$ phase factor.  This also determines the transformation properties of the electric field $e_{\br \br'}$ and vector potential $a_{\br \br'}$.  Due to the presence of the gauge structure, symmetries have nontrivial action on the spinon fields $f_{\br \alpha}$; the mathematical structure describing the realization of symmetries in such a situation is referred to as the \emph{projective symmetry group} (PSG).\cite{wen02}

If $S$ is a space group symmetry mapping $\br \to S(\br)$, then the above requirements dictate that the fermions transform as
\begin{equation}
S : f_{\br \alpha} \to \pi_{\br} f_{S(\br) \alpha} \text{,}
\end{equation}
where $\pi_{\br}$ is a gauge transformation depending on $S$; in the gauge chosen in Eq.~(\ref{eqn:effective-lattice-hamiltonian}) we can take $\pi_{\br} = \pm 1$.  Spin rotations send
$f_{\br \alpha} \to U_{\alpha \beta} f_{\br \beta}$, where $U$ is an ${\rm SU}(2)$ matrix.  Time reversal acts as follows:
\begin{equation}
{\cal T} : f_{\br \alpha} \to (i \sigma^2)_{\alpha \beta} f_{\br \beta} \text{.}
\end{equation}
These transformation laws also imply that the electric field and vector potential transform as scalars under translations, and as vectors under rotations and reflections.  Under time reversal, the electric field is even and the vector potential is odd.

Following the procedure in Appendix~\ref{app:symmetries}, we obtain the following transformations for the (real space) continuum field $\Psi(\br)$:
\begin{eqnarray}
T_{\ba_1} : \Psi &\to& (i \mu^2) \Psi \\
T_{\ba_2} : \Psi &\to& (i \mu^3) \Psi \\
{\cal R}_y : \Psi &\to& (i \tau^1) \exp \Big( \frac{i \pi}{2} \mu_{ry} \Big) \Psi \\
R_{\pi/3} : \Psi &\to& \exp \Big( \frac{i \pi}{6} \tau^3 \Big) \exp \Big( \frac{2 \pi i}{3} \mu_R \Big) \Psi \\
{\cal T} : \Psi &\to& (i\sigma^2)(i\tau^2)(-i \mu^2) \Psi  \label{eqn:tr-symm} \text{,}
\end{eqnarray}
where
\begin{eqnarray}
\mu_{ry} &=& - \frac{1}{\sqrt{2}} (\mu^1 + \mu^3 ) \\
\mu_R &=& \frac{1}{\sqrt{3}} ( \mu^1 + \mu^2 - \mu^3 ) \text{.}
\end{eqnarray}

\section{Many competing orders:  fermion bilinears and magnetic monopoles}
\label{sec:many-competing-orders}

\subsection{General discussion}
\label{sec:co-gendisc}

The focus of this paper is on observable properties of the algebraic spin liquid on the kagome lattice; the most dramatic such property is the unification of many competing orders within the ASL.\cite{hermele05}  This is manifested in the fact that correlations of a variety of superficially unrelated observables obey power law decay in space and time with the same 
critical exponent.  This unification is accomplished mathematically through the presence of an emergent ${\rm SU}(4)$ symmetry that contains the microscopic ${\rm SU}(2)$ spin rotation symmetry as a subgroup.  This means, for example, that magnetic order parameters can be rotated into nonmagnetic ones.  Furthermore, because the ASL is an interacting critical state, the critical exponents governing many physically interesting correlation functions are not 
equal to their mean field values.  In particular, for those observables we refer to as competing orders, the correlation functions decay more slowly than in mean field theory.  Such slow decay of a correlation function indicates that the system is closer to being ordered in a particular channel, and is likely to be observable in both experiments and numerical studies.

In the language of critical phenomena, the dominant long-distance correlations within the ASL, and hence the dominant ``competing order parameters,'' are given by finding the operators in the critical theory with smallest scaling dimension.  Suppose we are interested in the correlations of some microscopic observable $m(\br)$, which is some function of spin operators $\bS_{\br'}$ for $\br'$ near the lattice point $\br$.  Its long distance correlations can be understood from the following expression:
\begin{equation}
\label{eqn:micro-to-ft}
m(\br) \sim \sum_{i} c_i {\cal O}_i(\br) \text{.}
\end{equation}
On the right hand side of this expression are a set of operators in the field theory ${\cal O}_i$, also located at the spatial point $\br$.  The meaning of Eq.~(\ref{eqn:micro-to-ft}) is that the quantities on the left and right hand sides have the same correlations, at distances much greater than a short-distance cutoff.  For \emph{any} operator in the field theory, generically the coefficient $c_i$ will be nonzero if and only if ${\cal O}_i$ transforms identically to $m(\br)$ under microscopic symmetries (in this case, spin rotations, time reversal and space group operations).  Apart from this condition, the $c_i$ are nonuniversal.  This tells us, then, that the dominant long-distance correlations of $m(\br)$ are given by the ${\cal O}_i$ appearing in Eq.~(\ref{eqn:micro-to-ft}) with the smallest scaling dimension $\Delta$; so, for example, $\langle m(\br) m(0) \rangle \propto |\br|^{-2\Delta}$.

Which operators in the ASL critical theory have smallest scaling dimension?  In the large-$N_f$ limit, these are a set of fermion bilinears transforming as the adjoint of ${\rm SU}(N_f)$.  In the physical case of $N_f = 4$, these operators are
\begin{equation}
N^a = \Psi^\dagger \tau^3 T^a \Psi^{\vphantom\dagger} \text{,}
\end{equation}
where $a = 1, \dots, 15$ and the $T^a$ are the 15 generators of ${\rm SU}(4)$.  One can choose a basis for the generators where they are expressed in terms of the $\mu^i$ and $\sigma^i$ Pauli matrices:
\begin{equation}
T^a = \{ \sigma^i , \mu^i, \sigma^i \mu^j \} \text{.}
\end{equation}
In the large-$N_f$ limit, Rantner and Wen calculated the scaling dimension of one of the $N^a$ in the context of the ``staggered flux'' spin liquid on the square lattice.\cite{rantner02}  In Ref.~\onlinecite{hermele05} the ${\rm SU}(4)$ symmetry was emphasized, which implies that all the $N^a$ have the same scaling dimension.  The result of Ref.~\onlinecite{rantner02} is
\begin{equation}
\label{eqn:deltaN}
\Delta_N \equiv \operatorname{dim} N^a = 2 - \frac{32}{3 \pi^2 N_f} + {\cal O}(1/N_f^2) \text{.}
\end{equation}
Because the ${\rm U}(1)$ gauge interaction, which is stronger for smaller $N_f$, tends to bind together the $\Psi$ and $\Psi^\dagger$ fermions in $N^a$, and make the mode it creates propagate more coherently as opposed to decaying into its two constituent fermions, it is expected that $\Delta_N < 2$ for all values of $N_f$.

There are other operators in the ASL that may have even smaller scaling dimension than the $N^a$ when $N_f = 4$.  The most likely candidates are magnetic monopole operators -- these are topological disorder operators for the ${\rm U}(1)$ gauge field.  Because the ${\rm U}(1)$ gauge field is compact, its magnetic flux is quantized in multiples of $2\pi$, and monopole operators are those which insert integer multiples of $2\pi$ flux.  Therefore they carry nonzero integer charge under the ${\rm U}(1)_{{\rm flux}}$ symmetry, which emerges at the ASL fixed point and corresponds to the irrelevance of monopole operators there.\cite{hermele04}  It is important to emphasize that, for the ASL fixed point to be stable, all monopoles that are allowed as perturbations to the action by microscopic symmetries must be irrelevant.  The monopoles we will consider here carry nontrivial quantum numbers and are not allowed perturbations, so they may become relevant without destabilizing the ASL.

As is typical for topological disorder operators, it is difficult to construct monopole operators explicitly in terms of the fermions and gauge field.  However, they can be constructed by exploiting the state-operator correspondence from conformal field theory, and it has been found\cite{borokhov02} that monopoles with smallest scaling dimension at large-$N_f$ have unit magnetic charge and transform under the completely antisymmetric self-conjugate representation of the ${\rm SU}(N_f)$ flavor symmetry.  In the case $N_f = 4$ this representation is 6 dimensional.   Formally we can represent these six operators by $m^*_{\alpha \beta} = -m^*_{\beta \alpha}$, where $\alpha, \beta = 1,\dots,4$ are ${\rm SU}(4)$ indices.  There is another set of six monopoles with charge $-1$ and the same scaling dimension, $m_{\alpha \beta} = (m^*_{\alpha \beta})^\dagger$.  In the large-$N_f$ limit the scaling dimension of these operators was found to be
\begin{equation}
\Delta_m \equiv \operatorname{dim} m_{\alpha \beta} = c N_f + {\cal O}(1) \text{,}
\end{equation}
where $c \approx 0.265$.\cite{borokhov02}  If we na\"{\i}vely put $N_f = 4$ into the leading order large-$N_f$ result, we obtain $\Delta_m \approx 1.06$, compared to $\Delta_N \approx 1.73$ putting $N_f = 4$ into Eq.~(\ref{eqn:deltaN}).  So it may well be the case that the monopoles give the dominant long distance correlations for the physical case of $N_f = 4$.

Below, we shall work out the quantum numbers of $N^a$ and $m_{\alpha \beta}$ under microscopic symmetries, and determine the spin model observables to which they correspond.

\subsection{Fermion bilinears}
\label{sec:fermion-bilinears}

It is convenient to break the $N^a$ into three different classes of operators.  These are
\begin{eqnarray}
\bN^i_A &=& \Psi^\dagger \tau^3 \mu^i \bsigma \Psi^{\vphantom\dagger} \\
\bN_B &=& \Psi^\dagger \tau^3 \bsigma \Psi^{\vphantom\dagger} \\
N_C^i &=& \Psi^\dagger \tau^3 \mu^i \Psi^{\vphantom\dagger} \text{.}
\end{eqnarray}
Using the results of Sec.~\ref{sec:symmetries} we can determine how each of these observables transforms under all microscopic symmetries.  Clearly $\bN^i_A$ form a set of 3 spin triplets, and $\bN_B$ is also a spin triplet, while the $N^i_C$ are spin singlets. Under time reversal $\bN^i_A$ is odd, while $\bN_B$ and $N^i_C$ are even.  This implies that $\bN^i_A$ is the order parameter for a magnetically ordered state.  

For each of these operators, we can identify one or more microscopic operators that transform identically under all the symmetries of the kagome model.  Then, by Eq.~\ref{eqn:micro-to-ft}, the $N^a$ will contribute to the long-distance correlations of these microscopic operators.  Actually, rather than looking for operators, it is easier to look for ordering patterns (\emph{i.e.} states as opposed to operators) with the corresponding symmetry transformations.  To understand how to do this, let us consider the transformations of $\bN^i_A$.  If $S$ is a space group operation, then we have
\begin{equation}
\label{eqn:F1-matrices}
S : \bN^i_A \to U^{F_1}_{i j}(S) \bN^j_A \text{,}
\end{equation}
where $U^{F_1}(S)$ is an ${\rm O}(3)$ matrix.  These matrices form the $F_1$ irreducible representation of the space group (see Appendix~\ref{app:grouptheory} for more details).
 If we can find a set of magnetically ordered states for which $\langle \bS_{\br} \rangle$ transforms under the same representation of the space group, then $\bN^i_A$ is an order parameter for this state.  Furthermore, we can go on to construct microscopic operators that are also order parameters for this state, and identify correlation functions of these operators that obey power law decay with exponent $2 \Delta_N$.

The relevant details of the group theory and representation theory of the kagome space group are summarized in Appendix~\ref{app:grouptheory}.  As stated above, each component in spin space of $\bN^i_A$ transforms in the $F_1$ irreducible representation.  The $N^i_C$ also transform as $F_1$, that is $S : N^i_C \to U^{F_1}_{i j}(S) N^j_C$, and each component in spin space of $\bN_B$ transforms as the $A_2$ representation.

We have already established that the $\bN^i_A$ correspond to magnetically ordered states.  We are primarily interested in finding a combination of spin operators corresponding to each $\bN^i_A$, and for this purpose it is enough to consider collinear ordering patterns only.  We therefore focus on the $z$-component in spin space $(\bN^i_A)^z$.  

\begin{figure}
\includegraphics[width=3in]{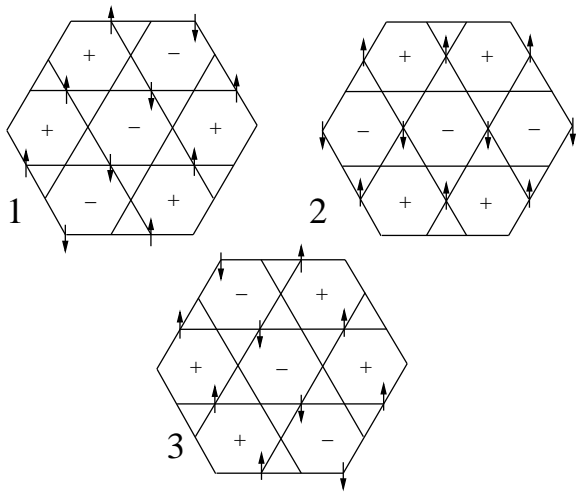}
\caption{Magnetic ordering patterns corresponding to $(\bN^i_A)^z$.  These can be thought of as stripes of up- and down-spin hexagons, which are labeled with ``+'' and ``-'', respectively.  A spin between two up-spin hexagons points up, and between two down-spin hexagons points down.  Spins between an up-spin and down-spin hexagon have no average moment.  The three patterns are labeled according to their crystal momenta, which lies at the three M-points of the Brillouin zone $M_i$, where $i = 1,2,3$.  The labeling of the M-points is as in Fig.~\ref{fig:unitcell-and-bz}.}
\label{fig:magorders}
\end{figure}

Our task is then to find three ordering patterns of kagome spins pointing along the $z$-axis in spin space.  One approach is simply to guess the right pattern, but it is possible to be more systematic.  We are interested in patterns of \emph{site} ordering on the kagome lattice that are invariant under translations by $2 \ba_1$ and $2 \ba_2$, since the $N^a$ are also invariant under such translations.  So we can consider all possible site ordering patterns on a cluster of $2 \times 2$ unit cells with 12 sites and periodic boundary conditions.  For each site of the cluster $\br$ we can associate a vector $| \br \rangle$, and the various site ordering patterns can be represented as real linear combinations of these states, where the coefficients of $| \br \rangle$ should be associated with $\langle S^z_{\br} \rangle$.  The space group acts on this vector space by $S | \br \rangle = | S(\br) \rangle$, where $S$ is a space group operation, so the vector space is a 12 dimensional representation of the space group.  This can be decomposed into irreducible representations using the character table of Appendix~\ref{app:grouptheory}.  The irreducible representations we find in this manner are a complete catalog of kagome site ordering patterns invariant under $T_{2 \ba_1}$ and $T_{2 \ba_2}$. We find that the $F_1$ representation of interest occurs precisely once in this catalog.   The corresponding magnetic ordering patterns can be read off from the basis vectors of this representation, and are shown in 
Fig.~\ref{fig:magorders}.  We stress that this is the \emph{unique} site-ordering pattern corresponding to $(\bN^i_A)^z$.

\begin{figure}
\includegraphics[width=2.5in]{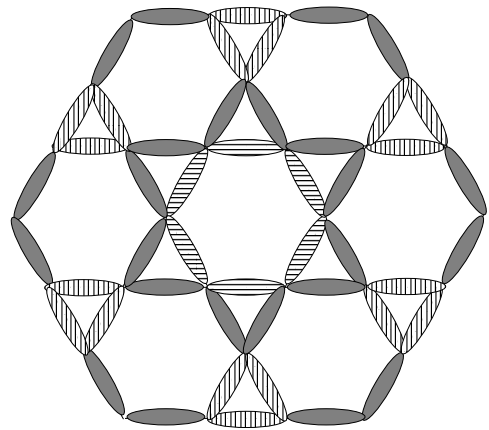}
\caption{Depiction of the Hastings valence bond solid state.  There are three inequivalent sets of bonds, these are distinguished by shading and vertical or horizontal hatching.  In previous treatments in the literature,\cite{hastings00, ran07} the shaded bonds were taken to have strong dimerization, and the vertically and horizontally hatched bonds to have weak (and equal) dimerization.  However, other choices are possible without changing the symmetry of the ground state.}
\label{fig:hastings-vbs}
\end{figure}

Next, we wish to find ordering patterns corresponding to $N^i_C$.  Because these operators are spin singlets and invariant under time reversal, it is natural to look for bond ordering patterns, with order parameters that can be built from $-\langle \bS_{\br} \cdot \bS_{\br'} \rangle$, where $\br$ and $\br'$ are nearest neighbors.  This observable measures the strength of singlet formation on a bond.  Following the same procedure as for the site ordering patterns, we note that the same unit cell contains 24 bonds; the resulting 24-dimensional vector space is decomposed into irreducible representations in Appendix~\ref{app:grouptheory}.  In this case, we find the $F_1$ representation occurs \emph{twice}, and leads to the patterns shown in Figs.~\ref{fig:f1a-bondorders},~\ref{fig:f1b-bondorders} and~\ref{fig:bondorders-superposed}.  Taken together with the ``uniform'' state, where $\langle \bS_{\br} \cdot \bS_{\br'} \rangle$ is the same on every bond, we can superpose the two patterns of Fig.~\ref{fig:bondorders-superposed} to form the Hastings VBS state, as shown in Fig.~\ref{fig:hastings-vbs}.  It has three inequivalent bonds in its unit cell, and this is precisely because it is built from a superposition of three bond ordering patterns belonging to distinct irreducible representations of the space group.  The fourfold degenerate Hastings state is not the only ground state that can be built from the bond ordering patterns obtained in Appendix~\ref{app:grouptheory}, but we restrict our attention to it for simplicity.  Furthermore, the ordering patterns obtained in Appendix~\ref{app:grouptheory} contain enough information to work out the contributions of $N^i_C$ to the bond-bond correlation function within the ASL.

\begin{figure}
\includegraphics[width=2.5in]{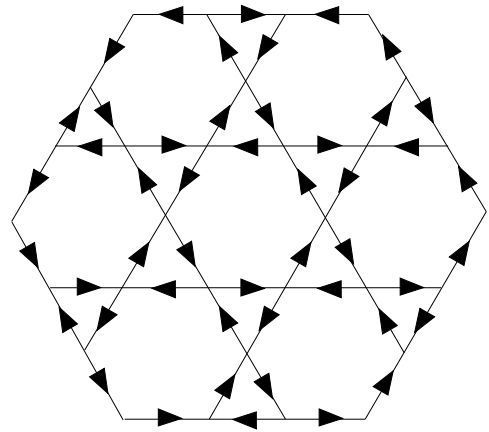}
\caption{Orientation of kagome lattice bonds used to define the vector spin chirality of a hexagon, $\bC_h(\br_h)$.}
\label{fig:orientation}
\end{figure}

Finally we turn to $\bN_B$.  This object is a spin triplet, but is even under time reversal; the simplest microscopic operator with these properties is the vector chirality $\bC_{\br \br'} = \bS_{\br} \times \bS_{\br'}$ defined on nearest-neighbor bonds.  Rather than searching for states as above,  in this case it is simpler to find a spin operator transforming identically to $\bN_B$.  We define an object that naturally resides on the hexagonal plaquettes of the kagome lattice, which we label with the position vectors $\br_h$:
\begin{equation}
\bC_h (\br_h) = \sum_{\langle \br \br' \rangle \in h} \bC_{\br \br'} \text{.}
\end{equation}
This can be interpreted as the vector spin chirality of a kagome hexagon.
Here, the sum is over the 6 bonds contained in the perimeter of the hexagon, following the orientation convention shown in Fig.~\ref{fig:orientation}.  More precisely, $\br$ is always taken to be the the ``first'' site on the bond according to Fig.~\ref{fig:orientation}, and $\br'$ the ``second'' site.  It is straightforward to check that $\bC_h (\br_h)$ has identical transformation properties as $\bN_B$ (once a suitable long-wavelength average is taken).  The observable $\bC_h$ is also related to the Dzyaloshinskii-Moriya (DM) interaction on the kagome lattice, and, in particular,
\begin{equation}
{\cal H}_{{\rm DM}} = D_z \sum_{\br_h} C^z_h(\br_h)
\end{equation}
is precisely the $D_z$-term allowed by the crystal symmetries in ZnCu$_3$(OH)$_6$Cl$_2$\cite{rigol07b}.  The effects of DM interaction on the ASL are discussed further in Sec.~\ref{sec:dm-interaction}.

\subsection{Magnetic monopoles}
\label{sec:monopoles}

Here, we work out the quantum numbers of the magnetic monopole $m^*_{\alpha \beta}$, which was formally defined in Sec.~\ref{sec:co-gendisc}, based on the results on Ref.~\onlinecite{borokhov02}.  Our analysis builds on results of Refs.~\onlinecite{alicea05a, alicea05b, alicea06, ryu07}, where monopole quantum numbers were worked out in different but closely related settings to the present one.  However, we adopt a different perspective; we believe this clarifies some of the issues involved, and we comment on this at the end of this section.  Our strategy is first to determine how each microscopic symmetry is embedded in the full symmetry group of the low-energy effective theory.  We do not have enough information to determine this completely, so several free parameters have to be introduced.  To proceed further, we argue that these free parameters must satisfy certain constraints, determined by relations among the generators of the space group.  After taking the constraints into account, we shall be left with three free parameters, and we make a conjecture that determines two of them, based on a physical argument and on calculations using the projected wavefunction.

First, it is useful to recall how $m^*_{\alpha \beta}$ transforms under the continuous symmetries of the low-energy ASL fixed point.  It is a scalar under Lorentz and continuous translation symmetry.  Under ${\rm SU}(4)$ rotations we have
\begin{equation}
m^*_{\alpha \beta} \to U_{\alpha \gamma} U_{\beta \delta} m^*_{\gamma \delta} \text{,}
\end{equation}
and under a ${\rm U}(1)_{{\rm flux}}$ rotation we have
\begin{eqnarray}
m^*_{\alpha \beta} &\to& e^{i \theta} m^*_{\alpha \beta} \\
m_{\alpha \beta} &\to& e^{- i \theta} m_{\alpha \beta} \text{.}
\end{eqnarray}

We now decompose the ${\rm SU}(4)$ symmetry into its ${\rm SU}(2) \times {\rm SU}(2)$ subgroup. The first ${\rm SU}(2)$ is simply spin rotation symmetry (generated by the $\sigma^i$), and the second consists of ``nodal'' rotations (generated by the $\mu^i$) that mix the two Dirac nodes but commute with spin rotations.  This decomposition is useful because it separates spin rotations from space group operations, which are realized in the ${\rm SU}(4)$ space as nodal ${\rm SU}(2)$ rotations.  We replace the ${\rm SU}(4)$ index $\alpha$ by a pair of ${\rm SU}(2)$ indices:  $\alpha \to (\sigma a)$.  Here, $\sigma = 1,2$ transforms under spin ${\rm SU}(2)$, and $a = 1,2$ under nodal ${\rm SU}(2)$.  Under an ${\rm SU}(2) \times {\rm SU}(2)$ rotation, we have
\begin{equation}
m^*_{(\sigma a) (\eta b)} \to U^S_{\sigma \sigma'} U^S_{\eta \eta'} U^N_{a c} U^N_{b d} m^*_{(\sigma' c) (\eta' d) } \text{,}
\end{equation}
where $U^S$ and $U^N$ are ${\rm SU}(2)$ matrices in the spin and nodal spaces, respectively.  We can also decompose the monopole operators according to their transformations under the ${\rm SU}(2) \times {\rm SU}(2)$ subgroup.  We have
\begin{eqnarray}
v^*_i = [ (i \sigma^2) \sigma^i ]_{\sigma \eta} (i \mu^2)_{a b} m^*_{(\sigma a) (\eta b)} \\
w^*_i = (i \sigma^2)_{\sigma \eta} [ (i \mu^2) \mu^i]_{a b} m^*_{(\sigma a) (\eta b)} \text{.}
\end{eqnarray}
Here, $v^*_i$ is a spin triplet and a nodal singlet, and $w^*_i$ is a spin singlet and nodal triplet.

We now wish to specify the action of microscopic symmetries on $v^*_i$ and $w^*_i$.  In doing this, we encounter two problems.  First, we determined the ${\rm SU}(4)$ rotation corresponding to each microscopic symmetry by working out the transformations of the fermion field $\Psi$.  This only determines the ${\rm SU}(4)$ rotation up to multiplication by the matrix $C_4 = \operatorname{diag}(i,i,i,i)$, which generates the $Z_4$ center of ${\rm SU}(4)$.  The reason for this ambiguity is that multiplication of $\Psi$ by $C_4$ is indistinguishable from a gauge transformation.  However, $m_{\alpha \beta}$ is odd under $C_4$, but is of course gauge invariant.  Second, each microscopic symmetry may come along with a rotation in the ${\rm U}(1)_{{\rm flux}}$ space.  Because $\Psi$ carries no ${\rm U}(1)_{{\rm flux}}$ charge, we have no information about this rotation.  Both of these ambiguities can be taken into account by multiplying $m_{\alpha \beta}$ by an undetermined phase factor associated with each symmetry operation.  We note that there are no such ambiguities associated with continuous spin rotations -- it can be shown that adding an additional ${\rm U}(1)$ phase factor to a spin rotation is not consistent with spin rotation symmetry.

We can now enumerate how the space group symmetries act on the monopoles:
\begin{eqnarray}
T_{\ba_1} : v_i^* &\to& e^{i \phi_{\ba_1}} v_i^* \\
T_{\ba_1} : w_i^* &\to& e^{i \phi_{\ba_1}} R_{i j}(T_{\ba_1}) w_j^* \\
{\cal R}_y : v_i^* &\to& e^{i \phi_{ry}} v_i \\
{\cal R}_y : w_i^* &\to& e^{i \phi_{ry}} R_{i j}({\cal R}_y) w_j \\
R_{\pi / 3} : v_i^* &\to& e^{i \phi_R} v_i^* \\
R_{\pi / 3} : w_i^* &\to& e^{i \phi_R} R_{i j}(R_{\pi/3}) w_j^* \text{.}
\end{eqnarray}
Note that the reflection symmetry sends $v_i^*$ to $v_i$, and similarly for $w_i^*$.  This is because reflections change the sign of the magnetic charge.  Here we have introduced ${\rm SO}(3)$ matrices describing the rotations in the nodal ${\rm SU}(2)$ space, which are obtained from the symmetry operations of Sec.~\ref{sec:symmetries} and are
\begin{eqnarray}
R(T_{\ba_1}) &=& \left( \begin{array}{ccc}
-1 & 0 & 0 \\
0 & 1 & 0 \\
0 & 0 & -1 \end{array} \right) \\
R({\cal R}_y) &=& \left( \begin{array}{ccc}
0 & 0 & 1 \\
0 & -1 & 0 \\
1 & 0 & 0 \end{array} \right)  \\
R(R_{\pi/3}) &=& \left( \begin{array}{ccc}
0 & 1 & 0 \\
0 & 0 & -1 \\
-1 & 0 & 0 \end{array} \right) \text{.}
\end{eqnarray}
We can immediately eliminate the phase factor $\phi_{ry}$ by redefining $v_i^* \to e^{i \phi_{ry}/2} v_i^*$, and similarly for $w_i^*$.  This does not affect the other symmetries, and we have
\begin{eqnarray}
{\cal R}_y :  v_i^* &\to&  v_i \\
{\cal R}_y : w_i^* &\to& R_{i j}({\cal R}_y) w_j \text{.}
\end{eqnarray}

We shall now determine the unknown phase factors $\phi_{\ba_1}$ and $\phi_R$ to the greatest extent possible, exploiting relations among the space group generators.  For example, we shall demand the relation $R_{\pi/3}^6 = 1$ holds when acting on monopole operators.  In general, such relations need only hold up to a phase when acting on quantum states.  However, in the present case, these relations hold with no additional phase factors when acting on any physical state in either the spin model or the effective lattice gauge theory Hilbert spaces.  Since nothing forbids the insertion of a single monopole in the lattice gauge theory, the relations must hold for the monopole operators with no extra phase factors.  The relation $R_{\pi/3}^6 = 1$ implies that $\phi_R = \pi n_R / 3$, where $n_R$ is an integer.  Next, the relation
\begin{equation}
R_{\pi/3} T_{\ba_1} R_{\pi/3} T_{\ba_1}^{-1} R^{-2}_{\pi/3} = T_{\ba_1}
\end{equation}
implies $\phi_{\ba_1} = 0$.  So we are left only with the free parameter $n_R = 0,1,2,3,4,5$.

We also need to consider time reversal.  We have
\begin{eqnarray}
{\cal T} : v^*_i &\to& t_v v_i \label{eqn:tr-mono1} \\
{\cal T} : w^*_i &\to& t_w w_i \label{eqn:tr-mono2} \text{,}
\end{eqnarray}
where $t_v, t_w = \pm 1$.  Note that time reversal changes the sign of magnetic charge, so it takes monopoles to antimonopoles.  Furthermore, time reversal commutes with the ${\rm SU}(2) \times {\rm SU}(2)$ subgroup of ${\rm SU}(4)$ [see Eq.~(\ref{eqn:tr-symm})], so it must take the spin triplet $v^*_i$ to the spin triplet $v_i$, and similarly for the nodal triplets $w^*_i$ and $w_i$. It is not consistent for $t_v$ and $t_w$ to be arbitrary phases, given that ${\cal T}^2 = 1$, which holds for all physical states of the spin model and of the effective lattice gauge theory.  The relations of  Eq.~(\ref{eqn:tr-mono1}) and Eq.~(\ref{eqn:tr-mono2}) are the most general transformations consistent with these considerations.

We proceed by using some physical input to conjecture the likely values of $n_R$ and $t_v$ -- the resulting conjecture is supported by the numerical results on the projected wavefunction (see Sec.~\ref{sec:spin-correlation}).  The source of physical input is the following puzzle about the kagome ASL:  in the large-$S$ limit of the kagome antiferromagnet, there are an infinite number of nearly degenerate magnetically ordered ground states.  These consist of all states where the vector sum of classical spins on every triangle is zero -- among these states, those with coplanar spin configurations are selected by the zero point energy of spin wave fluctuations about the ground state.  Finally, among the coplanar states, the $\sqrt{3} \times \sqrt{3}$ state is selected by anharmonic fluctuations.\cite{chubukov92,chandra93,henley95}  Up to this point in our analysis, we have not found signatures of any of these classical low energy states within the ASL.  (The ordered states of Fig.~\ref{fig:magorders} have distinct symmetry from the classical ground states lying at the M points, which transform in the $F_3$ representation of the space group.)  It is possible that no sign of the large-$S$ physics survives down to $S = 1/2$, but it would perhaps be more natural if at least a hint of it remained, especially given that the ASL has no spin gap.  

\begin{figure}
\includegraphics[width=2in]{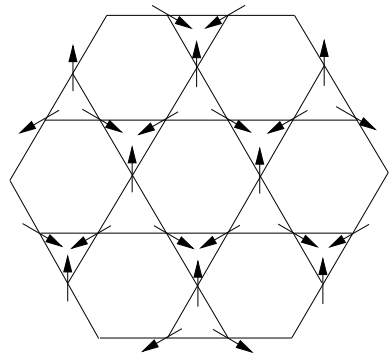}
\caption{The $\bq = 0$ magnetically ordered state.}
\label{fig:qzero-state}
\end{figure}

This leads us to the conjecture that $v_i$ will be the order parameter for one of these classical ground states.  We know from the analysis above that $v_i$ carries zero crystal momentum, and the unique classical low-energy state with this property is the so-called $\bq = 0$ state, shown in Fig.~\ref{fig:qzero-state}.  It is the classical ground state for ferromagnetic third-neighbor exchange $J_3 < 0$.\cite{chubukov92}  It turns out that if we choose $n_R = 2$ and $t_v = -1$, then $v_i$ is an order parameter for the $\bq = 0$ state.  This can be seen by following the construction of the $\bq = 0$ state order parameter in Appendix~\ref{app:qzero-state}.  Based on these considerations, from now on we shall fix $n_R = 2$ and $t_v = -1$.  This choice is supported by the presence of substantial $\bq = 0$ spin correlations in the projected wavefunction.

The parameter $t_w = \pm 1$ still remains to be determined.  The Hermitian operators
$w^+_i = w_i + w^*_i$ and $w^-_i = i ( w_i - w^*_i)$ are spin singlets, with crystal momenta lying at the M-points.  If $t_w = 1$, they are even under time reversal, and then likely correspond to bond ordering patterns.  If $t_w = -1$, they are odd under time reversal, and correspond to ordering patterns in the scalar spin chirality $\bS_1 \cdot (\bS_2 \times \bS_3)$, where the spins can be taken on any three distinct nearby lattice sites.  Since the bond-bond correlations in the projected wavefunction are apparently dominated by Fourier components near $\bq = 0$, we are led to speculate that  $t_w = -1$ and the monopoles correspond to spin chirality order.

We now contrast the approach taken here with that of Refs.~\onlinecite{alicea05a, alicea05b, alicea06, ryu07}, where symmetry transformations of monopole operators were obtained by construction of mean-field ground state wavefunctions with a background $\pm 2\pi$ flux -- this is the quantum state one obtains after insertion of a monopole operator.  The correspondence of the resulting states to scaling operators of the low-energy critical theory, in which one is ultimately interested, is not clear, and this issue makes it difficult to interpret the results of Refs.~\onlinecite{alicea05a, alicea05b, alicea06, ryu07}.  However, our approach here, which focuses directly on the scaling operator $m^*_{\alpha \beta}$, is formally equivalent to the analysis of Refs.~\onlinecite{alicea05a, alicea05b, alicea06, ryu07}, and puts it on firm ground.  On the other hand, Refs.~\onlinecite{alicea05a, alicea05b, alicea06, ryu07} obtained additional information on symmetries by carrying out an adiabatic insertion of $2\pi$ flux numerically.  In our opinion, because the ASL has gapless excitations, the correspondence between this procedure and insertion of the scaling operator $m^*_{\alpha \beta}$ is not established.

\section{Detection of competing orders}
\label{sec:detection}

The competing orders arising in the kagome ASL can be detected by a variety of experimental probes.  Here, we outline some predictions that we hope will be tested in future experiments on ZnCu$_3$(OH)$_6$Cl$_2$.  In this section we focus on the case of an ideal system free of perturbations such as impurities and Dzyaloshinskii-Moriya interaction.  We postpone discussion of these issues to the following section, and also Sec.~\ref{sec:discussion}.

Two critical exponents enter into this discussion.  The first is $\eta_N = 2 \Delta_N - 1$, which characterizes the correlations of the $N^a$ fermion bilinears, and hence the M-point magnetic order, the Hastings VBS order, and the hexagon vector chirality $\bC_h(\br_h)$.  The second, $\eta_m = 2 \Delta_m - 1$, characterizes the $m_{\alpha \beta}$ monopole operators, and hence the correlations of $\bq = 0$ magnetic order.

The magnetic competing orders can be detected via neutron scattering and NMR.  
In neutron scattering, the structure factor will exhibit critical scaling behavior near the $\Gamma$ and $M$ points in reciprocal space, with the scaling form
\begin{equation}
\chi''(\bq + \bQ, \omega) = \frac{c}{|\omega|^{2 - \eta}} F_{\bQ}\Big( \frac{\hbar \omega}{k_B T}, \frac{\omega}{v_F | \bq - \bQ|} \Big) \text{,}
\label{eqn:neutron-scaling}
\end{equation}
where $c$ is a nonuniversal constant prefactor, and $|\bq|$ is much smaller than the Brillouin zone size.  $\eta = \eta_N$ if $\bQ$ lies at one of the M points, and $\eta = \eta_m$ if $\bQ = 0$.  There are two different universal scaling functions $F_{\bQ}(x,y)$, again depending on whether $\bQ$ lies at the $\Gamma$ point or one of the M points.

Based on Eq.~(\ref{eqn:neutron-scaling}), the NMR relaxation rate is predicted to be
\begin{equation}
\frac{1}{T_1} \propto T^{\eta} \text{.}
\end{equation}
Here, $\eta$ depends on the nuclear site considered -- Cu, O and Cl sites are all sensitive to the M-point magnetic order, but only Cu and O are sensitive to $\bq = 0$ order.  Therefore we expect
\begin{eqnarray}
\eta_{{\rm Cu}} &=& \eta_{{\rm O}} = \operatorname{min}(\eta_N, \eta_m) \\
\eta_{{\rm Cl}} &=& \eta_N \text{.}
\end{eqnarray}

The VBS competing order can be detected via its coupling to phonons.  We consider, for simplicity, a single optical phonon that couples to one of the patterns of VBS order shown in Figs.~\ref{fig:f1a-bondorders} and~\ref{fig:f1b-bondorders}, and hence to some linear combination of the $N^i_C$ fermion bilinears of the ASL.  Using inelastic X-ray or neutron scattering, the lineshape of this phonon can be measured.  The lineshape is determined by $D(\omega)$, the retarded Green's function for the phonon mode's normal coordinate.  We treat the phonon mode using the RPA of Ref.~\onlinecite{cross79}, which neglects the influence of the gapped optical phonon on the gapless spin system, and has been successfully applied to quasi-one-dimensional magnets, for temperatures above the spin-Peierls transition.\cite{abel07}  We then have, for the phonon spectral function $A(\omega) = - (1/\pi) \operatorname{Im} D(\omega)$.
\begin{equation}
A(\omega) =  \frac{- g^2 \chi''_{{\rm VBS}}(\omega, T) / \pi}
{\big[ \omega^2 - \Omega^2_0 - g^2 \chi'_{{\rm VBS}}(\omega, T) \big]^2 + \big[ g^2 \chi''_{{\rm VBS}}(\omega, T) \big]^2} \text{.}
\end{equation}
Here, $g$ characterizes the spin-phonon coupling, $\Omega_0$ is the bare phonon frequency, and $\chi'_{{\rm VBS}}$ and $\chi''_{{\rm VBS}}$ are the real and imaginary parts, respectively, of the retarded Green's function $\chi_{{\rm VBS}}$ of the VBS order parameter.  We have the scaling form
\begin{equation}
\chi_{{\rm VBS}}(\omega, T) = \frac{c}{|\omega|^{2 - \eta_N}} F_{{\rm VBS}} \Big( \frac{\hbar \omega}{k_B T} \Big) \text{,}
\end{equation}
where $c$ is a nonuniversal prefactor.  Also, $F_{{\rm VBS}}$ is related to the scaling function appearing in Eq.~(\ref{eqn:neutron-scaling}) by $\operatorname{Im} F_{{\rm VBS}}(x) = F_{\bQ}(x, 0)$, for $\bQ$ lying at one of the M points.  It may be possible to test this scaling form by measuring the $T$-dependence of $A(\omega)$.

\section{Dzyaloshinskii-Moriya interaction}
\label{sec:dm-interaction}

We now consider the effect of a small DM interaction, $D_z, D_p \ll J$.  First, we must understand which new perturbations to the ASL fixed point are allowed by the now reduced microscopic symmetry.  While spin rotation symmetry is completely broken, the space group symmetry remains unchanged.  However, lattice reflections and rotations must now be accompanied by appropriate operations in spin space.  In the three-dimensional Herbertsmithite structure, the reflection ${\cal R}_y$ is realized as a $\pi$-rotation about the $x$-axis (passing through the center of a hexagon), along with a corresponding rotation in spin space.  The rotation $R_{\pi/3}$ is realized by first making the same $\pi$-rotation about the $x$-axis, followed by mirror reflection through the plane with normal $\ba_2 - \ba_1$ (intersecting the center of the same hexagon).  In the continuum, the resulting modified symmetry operations are
\begin{eqnarray}
{\cal R}'_y : \Psi &\to& (i \sigma^1) (i \tau^1) e^{i \pi \mu_{ry} / 2}\Psi \\
R'_{\pi/3} : \Psi &\to&  e^{i \pi \sigma^3 / 3} e^{i \pi \tau^3 / 6} e^{2 \pi i \mu_R / 3} \Psi \text{.}
\end{eqnarray}

Using these operations, combined with translations and time reversal (which are unchanged), it can be shown that the only fermion bilinear allowed by symmetry is $N^z_B$.  Furthermore, the $q = 1$ monopole operators $m^*_{\alpha \beta}$ are still forbidden by symmetry -- the spin triplet monopoles $v_i$ are odd under time reversal, and the spin singlet monopoles $w_i$ carry nonzero crystal momentum.  Therefore we expect $N^z_B$ to be the most relevant perturbation to the ASL generated by the DM terms.

We now consider adding the term
\begin{equation}
\label{eqn:dm-perturbation}
{\cal L}_{{\rm DM}} = m N^z_B 
\end{equation}
to the ASL fixed point.  We wish to understand how $m$ depends on $D_z$ and $D_p$.  As noted in Sec.~\ref{sec:fermion-bilinears}, the $D_z$-term and $N^z_B$ are symmetry-equivalent, so $m$ will contain a term linear in $D_z$.  We now give an argument that $m$ contains no term linear in $D_p$.  Consider a microscopic Hamiltonian with given values of $D_p$ and $D_z$.  Upon some amount of coarse-graining, we obtain a continuum field theory with a given value of $m$.  Now we make a $\pi$-rotation in spin space about the $z$-axis.  This sends $D_p \to - D_p$, but otherwise leaves the microscopic Hamiltonian unchanged.  Various operators in the continuum field theory will be affected by this operation, but the value of $m$ remains the same.  We have shown that $m$ does not depend on the sign of $D_p$.  Therefore, up to second order in $D_z$ and $D_p$,
\begin{equation}
m = c_{z 1} D_z + c_{z 2} \frac{D_z^2}{J} + c_{p 2} \frac{D_p^2}{J} \text{,}
\end{equation}
where $c_{z 1}$, $c_{z 2}$ and $c_{p 2}$ are dimensionless coefficients.

Ignoring coupling between the gauge field and the fermions, the effect of the perturbation Eq.~(\ref{eqn:dm-perturbation}) is to open a spin gap.  However, coupling to the gauge field plays a surprising and important role.
As the fermions are now massive, magnetic monopoles will condense, leading to confinement.\cite{polyakov77}  Na\"{\i}vely this leads to a fully gapped spectrum, but we argue below that, in this case it leads instead to XY magnetic order.

Let us first imagine $D_z \neq 0$, but $D_p = 0$, so that we have ${\rm U}(1)$ spin rotation symmetry about the $z$-axis in spin space.  The mass term ${\cal L}_{{\rm DM}}$ will be induced, and we observe that this term can be written
\begin{equation}
{\cal L}_{{\rm DM}} = m \sum_{a = \pm} \Big[ \psi^\dagger_{\uparrow a} \tau^3 \psi^{\vphantom\dagger}_{\uparrow a}
- \psi^\dagger_{\downarrow a} \tau^3 \psi^{\vphantom\dagger}_{\downarrow a} \Big] \text{.}
\end{equation}
That is, this term has a mass $m$ for the up-spin fermions, and a mass of opposite sign $-m$ for the down-spin fermions.  Ignoring coupling to the gauge field for the moment, such a mass term leads to a $\nu = 1$ quantum Hall effect (QHE) for the up-spin fermions, and a $\nu = -1$ QHE for spin-down fermions.\cite{haldane88}

Now, imagine we adiabatically insert a localized $+2\pi$ flux of the gauge field.  The QHE implies that a single extra spin-up spinon will be induced along with the gauge flux, while one spin-down spinon will be depleted.  This is equivalent to a spin flip operation, and we see that insertion of a monopole is accompanied by a spin flip.  Furthermore, because the fermions are gapped, the gauge field dynamics is controlled by the Maxwell term ${\cal L}_{{\rm Maxwell}} = \frac{1}{2 e^2} \sum_{\mu} (\epsilon_{\mu \nu \lambda} \partial_{\nu} a_{\lambda})^2$, and therefore insertion of a monopole only costs finite action in the imaginary time functional integral.  This implies that the monopole propagator is long-ranged -- that is, if $m^*(\br)$ is a monopole creation operator, then $\langle m^*(\br) m(0) \rangle$ approaches a constant as $|\br| \to \infty$.  Because monopole insertion is accompanied by a spin flip, in the ground state we will have $\langle S^+ \rangle \neq 0$, which corresponds to XY magnetic ordering.  Furthermore, within the resulting ordered state, the photon of the gauge field is expected to correspond to the XY Goldstone mode.  We note that a similar situation arises, in the absence of DM interaction, when a Zeeman magnetic field is applied.\cite{ran07b}

In Sec.~\ref{sec:monopoles}, we argued that some of the monopole operators in the ASL correspond to the $\bq = 0$ ordered state.  We therefore expect that the XY order induced in the presence of the $D_z$ term is in the $\bq = 0$ pattern, and therefore that the   ASL is unstable toward  $\bq=0$ magnetic ordering in the presence of $D_z$.  We remark that, once $D_p \neq 0$, the ${\rm U}(1)$ spin rotation symmetry is lost and there is no longer an XY Goldstone mode.  On the other hand, time-reversal is still a good symmetry of the Hamiltonian, but is broken in the $\bq = 0$ state.  So, for more general DM interaction ($D_p \neq 0$), we expect the ASL is unstable to a time-reversal broken ground state.

\section{Analysis of projected wavefunction}
\label{sec:analysis-pw}

We now turn to an analysis of the Gutzwiller projected variational wavefunction for the ASL.  To obtain this wavefunction, one begins with $|\psi^0_{{\rm MFT}} \rangle$, the mean-field ground state of the Hamiltonian Eq.~(\ref{eqn:mft-hamiltonian}).  The trial wavefunction is obtained from this by writing
\begin{equation}
| \psi^0_{{\rm prj}} \rangle = {\cal P}_G | \psi^0_{{\rm MFT}} \rangle \text{,}
\end{equation}
where the Gutzwiller projection operator,
\begin{equation}
{\cal P}_G = \prod_{\br} n_{\br} (2 - n_{\br}) \text{,}
\end{equation}
with $n_{\br} = f^\dagger_{\br \alpha} f^{\vphantom\dagger}_{\br \alpha}$, enforces the single occupancy constraint.  Properties of these wavefunctions can be calculated numerically using a standard Monte Carlo technique.\cite{gros89}

In Ref.~\onlinecite{ran07}, we found that $|\psi^0_{{\rm prj}} \rangle$ has the lowest energy among a class of spin liquid wavefunctions.  Furthermore, the energy is very low, less than $1\%$ above an extrapolation of the exact ground state energy to the thermodynamic limit;\cite{waldtmann98} this is achieved without tuning any continuous variational parameters to minimize the energy. By continuously deforming $|\psi^0_{{\rm MFT}} \rangle$ to include a small VBS ordering of the Hastings type and measuring the resulting variational energy, we argued that ASL is locally stable to this type of ordering.  Here, we discuss some of the properties of this wavefunction.  First, we construct variational excited states to estimate the velocity $v_F$ characteristic of the ASL.  Next, we proceed to discuss the spin-spin and bond-bond correlation functions of the wavefunction, and connect them to our understanding of the low-energy effective theory.

We consider finite systems with periodic boundary conditions for the physical spin operators, so that
\begin{equation}
\bS(\br + L_1 \ba_1) = \bS(\br + L_2 \ba_2) = \bS(\br) \text{.}
\end{equation}
Such a system has $3 L_1 L_2$ sites.  This still allows the consideration of twisted boundary conditions for the fermions -- we consider both periodic and antiperiodic fermion boundary conditions.
A technical point that plays a role in some of our analysis is that, depending on the system size, it is not always possible to construct a projected wavefunction that is fully symmetric under the kagome space group.  The details of this subtlety are given in Appendix~\ref{app:construct-prj}.  There we show how to construct symmetric wavefunctions for $L_1 = L_2 = 4 n$.  The technique of Ref.~\onlinecite{gros89} is easily extended to calculate with these wavefunctions, which are superpositions of three projected single Slater determinants, and all the correlation function results shown here are for these fully symmetric wavefunctions.

\subsection{Fermi velocity}
\label{sec:fermi-velocity}

\begin{figure}
\includegraphics[width=0.4\textwidth]{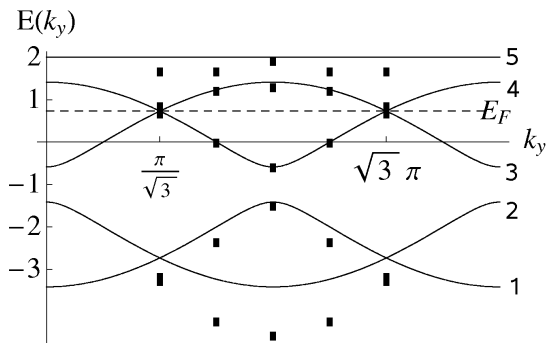}
\caption{Plot of the mean-field band structure (solid lines) and projected band structure (solid rectangles), obtained as discussed in the text. The energy is in units of $\chi J$. The band structure is plotted along the line from $-$M$_2$ to M$_2$ in the Brillouin zone.  The projected band structure can be fit by the mean-field band structure, choosing  $\chi=0.40 \pm 0.04$.  The band indices are shown on the right.}
\label{fig:prj_bs_plot}
\end{figure}

An important parameter in the algebraic spin liquid is the ``Fermi velocity'' $v_F$.  Physically, $v_F$ is defined as the ratio of energy and momentum for \emph{any} low energy excitation -- because the ASL is Lorentz invariant at low energies, all excitations propagate with the same velocity.  In principle, $v_F$ can be directly measured via inelastic neutron scattering, by tracking the leading edge of the scattering continuum near the $\Gamma$ point and the M points, where the spin gap is predicted to vanish.  Furthermore, it appears in the coefficients of specific heat and magnetic susceptibility, and in various critical scaling functions that may be measurable.

To estimate $v_F$, we need to generate low energy excitations of the ASL. A natural way to do this is to make a particle-hole excitation of the mean-field state, then act with the projection operator:
\begin{equation}
\vert \psi (\bk, \bq, i, j) \rangle = {\cal P}_G f_{\bk i \uparrow}^{\dagger} f^{\vphantom\dagger}_{\bq j \uparrow}\vert \psi^0_{{\rm MFT}} \rangle \text{.}
\end{equation}
One can calculate the variational energy of the excited state $\vert \psi (\bk, \bq, i, j) \rangle$ (later referred to as ``projected particle-hole excitation'') with respect to the ground state as a function of $\bk$, $\bq$, and the band indices $i, j$. We label the bands by indices $1,2, \dots, 5$ as shown in Fig.\ref{fig:prj_bs_plot}. Note that band 5 is doubly degenerate.

The spectrum of projected particle-hole excitations can
be translated into a projected band structure. This is
simplest if there is a single-particle (or single-hole) ex-
citation that costs zero energy in the mean-field band
structure. In this case, if the hole in the particle-hole ex-
citation has zero energy, we can interpret the energy as
that of the particle, and vice-versa. This situation can be
arranged by choosing the boundary condition such that
there are fermions precisely at the nodal points in mo-
mentum space. Particles and holes created at the nodal
points (in bands 3 and 4) have zero energy in mean-field
theory. We study the projected particle-hole excitations
for $L_1 = L_2 = 8$ with periodic-periodic boundary condi-
tions, which ensures that we have nodal fermions. (This
boundary condition, on the other hand, has a higher
ground state energy than that of periodic-antiperiodic
boundary conditions, in which case there are no nodal
fermions.)

To extract the projected band structure, we fix the hole at one node in band 3, then scan the momentum and band label of the particle in bands 4 and 5 and find the energies of this projected state $E(\vert \psi (\bk, \bq, i, j) \rangle)$. This is essentially a particle excitation and we plot $E(\vert \psi (\bk, \bq, i, j) \rangle)-E_{GS}$ ($E_{GS}$ is the projected ground state energy) in Fig.~\ref{fig:prj_bs_plot} with respect to the fermi energy. Similarly we fix the particle at one node in band-4, then scan the momentum and band label of the hole in bands 1, 2 and 3. This is a hole excitation and we plot $E_{GS}-E(\vert \psi (\bk, \bq, i, j) \rangle)$ instead.

It should first be noted that when both $\bk$ and $\bq$ lie right
on the nodal points in bands 3 and 4, the excitation en-
ergy is very small compared to the bandwidth, even if the particle and hole are at two
different nodes. For example when the particle and hole are at the same node, we find the excitation energy relative to the ground state to be $0.03(1)$J; when the particle and hole are at opposite node, we find the excitation energy to be $0.01(1)$J.
This
is consistent with the mean-field band structure, where
such excitations cost zero energy. We can fit the low energy projected particle-hole excitations in bands 3 and
4 (\emph{i.e.} the low-energy excitations) with the mean-field
band structure by tuning the hopping parameter $\chi$. (Recall that $\chi = t / J$ is the mean-field hopping in units of $J$.)  We
find $\chi = 0.40 \pm 0.04$ (the mean-field value $\chi_{mean}=0.221$ in Hastings' study\cite{hastings00}).  The parameter $\chi$ determines the fermi velocity $v_F=\frac{a\chi J}{\sqrt{2}\hbar}$, where $a$ is the lattice spacing and $J / k_B \approx 200 {\rm K}$ for herbersmithite. Therefore, for herbertsmithite, we estimate $v_F\approx 5.0\times 10^3$m/s.

\subsection{Spin-spin correlation function}
\label{sec:spin-correlation}

\begin{figure}
\includegraphics[width=3.0in]{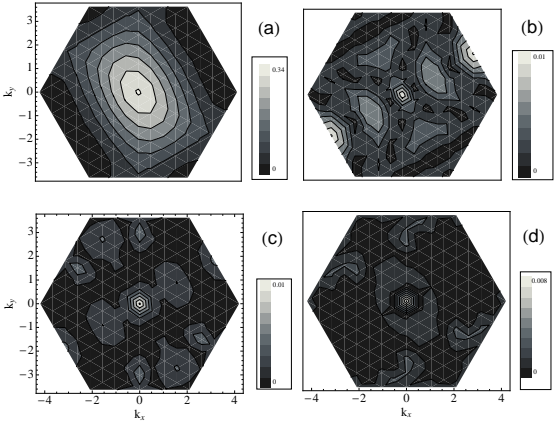}
\caption{Contour plots of $|S_{00}(\bq, d_{{\rm min}})|$ for (a) $d_{{\rm min}} = 0$, (b) $d_{{\rm min}} = \sqrt{3}$,
(c) $d_{{\rm min}} = 2$, and (d) $d_{{\rm min}} = \sqrt{7}$.  The scale is shown to the right of each plot.  The statistical error is roughly $1.5 \times 10^{-4}$, much less than the $\bq = 0$ peak height for all values of $d_{{\rm min}}$ shown.  }
\label{fig:scplots}
\end{figure}

\begin{figure}
\includegraphics[width=2.8in]{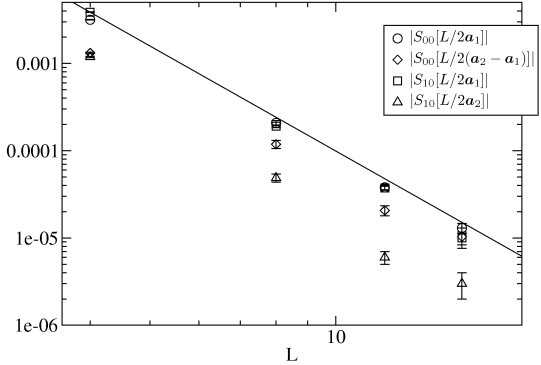}
\caption{Real space spin correlations measured at a distance of half the system size, for $L = 4,8,12,16$.  The straight line is a plot of the function $1/L^4$.}
\label{fig:sc-realspace}
\end{figure}

We consider the spin-spin correlation function defined by
\begin{equation}
S_{i j}(\bR) = \frac{1}{3} \langle \bS_{\bR i} \cdot \bS_{0 j} \rangle \text{,}
\end{equation}
and its Fourier transform $S_{i j}(\bq) = \sum_{\bR} e^{- i \bq \cdot \bR} S_{i j}(\bR)$.  We are primarily interested in the spin correlations at long distances; in order to better understand these, it is convenient to define the following object:
\begin{equation}
S_{i j} (\bq, d_{{\rm min}} ) = \sum_{ |\br(\bR,i) - \br(0,j)| \geq d_{{\rm min}} } e^{- i \bq \cdot \bR} S_{i j}(\bR) \text{.}
\end{equation}
The distance $| \br(\bR,i) - \br(0,j) |$ above is understood to mean the shortest distance between the lattice points $\br(\bR,i)$ and $\br(0,j)$, accounting for periodic boundary conditions.
$S_{i j} (\bq, d_{{\rm min}} )$ is the Fourier transform of the spin correlation function, with all pairs of sites with separation below $d_{{\rm min}}$ removed.

In Fig.~\ref{fig:scplots} we have plotted $|S_{0 0}(\bq, d_{{\rm min}})|$ for $L_1 = L_2 = 12$ at four different values of $d_{{\rm min}}$.  The $\bq = 0$ correlations dominate as $d_{{\rm min}}$ is increased; this trend continues for larger values of $d_{{\rm min}}$.  The same is true for $|S_{1 0}(\bq, d_{{\rm min}})|$ (not shown).  Furthermore, we find that $S_{0 0}(0, d_{{\rm min}}) > 0$ and $S_{1 0}(0, d_{{\rm min}}) < 0$ for all values of $d_{{\rm min}}$ in the $L_1 = L_2 = 12$ system (although, for the very largest values of $d_{{\rm min}}$, the result may be dominated by statistical error).  This suggests that the long-distance $\bq = 0$ correlations are dominated by the $\bq = 0$ pattern of magnetic order shown in Fig.~\ref{fig:qzero-state}.

One expects that the $\bq = 0$ spin correlations decay as a power law; the exponent of the power law decay can be estimated from the behavior of $S_{i j}(\bR)$ in real space.  In particular, for a system size $L = L_1 = L_2$, we consider the behavior of $S_{i j}(\bR)$ measured at a distance $|\bR| = L/2$.  The $00$ component of $S_{i j}(\bR)$ is plotted for the two inequivalent directions $\bR = (L/2) \ba_1, (L/2) (\ba_2 - \ba_1)$, and the $01$ component is plotted for $\bR = (L/2)\ba_1, (L/2)\ba_2$.  In all four cases,
the data are consistent with
\begin{equation}
S_{i j}(\bR) \propto \frac{1}{L^4} \text{.}
\end{equation}
This behavior leads to an estimate $\Delta_m \approx 2$ for the monopole scaling dimension (Fig.~\ref{fig:sc-realspace}).

It is perhaps surprising that the M-point magnetic orders depicted in Fig.~\ref{fig:magorders} do not appear to contribute significantly to the spin correlations of the projected wavefunction.  In particular, if the estimate $\Delta_m \approx 2$ is accurate, we expect the M-point correlations to decay more slowly, dominating at long distances.  This follows from the expectation $\Delta_N < 2$.\cite{rantner02}  We wish to make two comments on this result.  First, we emphasize again that the projected wavefunction may not give correct values for critical exponents of the ASL.  Second, the M-point correlations in the wavefunction may indeed decay more slowly than the $\bq = 0$ correlations, but with a much smaller coefficient.

\subsection{Bond-bond correlation function}
\label{sec:bond-correlation}

\begin{figure}
 \includegraphics[width=0.3\textwidth]{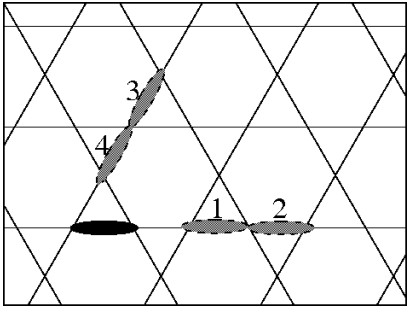}
\caption{We illustrate the relative bond positions and orientations that lead to the four inequivalent components of the bond-bond correlation function.  ${\cal C}^i(\bR)$ is given by the expectation value of the product of the dimerization of the black bond, with the dimerization of the gray-shaded bond at position $i$.  For example, the gray shaded bond at position 1 gives ${\cal C}^1(\ba_1)$, and other values of ${\cal C}^1(\bR)$ are obtained by translating the gray bond by a lattice vector.}\label{bond_bond_corr}
\end{figure}

\begin{figure}
 \includegraphics[width=3in]{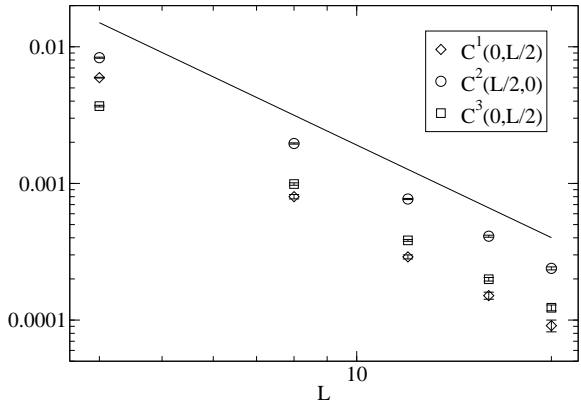}
\caption{Plot of bond-bond correlation function at a distance of half the system size. The straight line is a plot of the function $L^{-2.25}$.}\label{bond_edge_plot}
\end{figure}

The bond-bond correlation function has four inequivalent components, depending on the relative position and orientation of the two bonds involved.  These are illustrated in Fig.~\ref{bond_bond_corr}, and the corresponding components of the correlation function are
\begin{eqnarray}
{\cal {C}}^1(\bR) &=& \langle ( \bS_{\bR,0} \cdot \bS_{\bR,1}) (\bS_{\boldsymbol{0},0} \cdot \bS_{\boldsymbol{0},1} ) \rangle - \bar{B}^2 \\
{\cal C}^2 (\bR) &=&  \langle ( \bS_{\bR,1} \cdot \bS_{\bR + \ba_1,0}) (\bS_{\boldsymbol{0},0} \cdot \bS_{\boldsymbol{0},1} ) \rangle - \bar{B}^2  \\
{\cal C}^3 (\bR) &=&  \langle ( \bS_{\bR,0} \cdot \bS_{\bR, 2}) (\bS_{\boldsymbol{0},0} \cdot \bS_{\boldsymbol{0},1} ) \rangle - \bar{B}^2  \\
{\cal C}^4 (\bR) &=&  \langle ( \bS_{\bR,2} \cdot \bS_{\bR+\ba_2, 0}) (\bS_{\boldsymbol{0},0} \cdot \bS_{\boldsymbol{0},1} ) \rangle - \bar{B}^2  
\end{eqnarray}
where $\bar{B} \equiv \langle \bS_{\boldsymbol{0},0} \cdot \bS_{\boldsymbol{0},1} \rangle$ is the average dimerization, and the subtraction of $\bar{B}^2$ ensures that 
that the correlation function decays to zero at infinity.

Because computation of bond-bond correlation is expensive for larger systems, we only compute some particular values at a distance of half of the system size.  For $L_1 = L_2 = L$ we consider either $\bR = (L/2)\ba_1$, or $\bR = (L/2)\ba_2$. Table~\ref{bond_edge_data} lists the five sets of data that we obtained. We note that, in the second and third rows of Table~\ref{bond_edge_data}, the correlation function changes sign (for a relatively small value of $L$) as $L$ increases.  This signals that these data sets may be further from the scaling limit than those that do change sign.  Therefore, in Fig.~\ref{bond_edge_plot} we only plot the first, fourth and fifth sets of data, which are consistent with power-law decay. We estimate the power law to be
\begin{equation}
\label{eqn:vbs-powerlaw}
{\cal C}^i(\bR) \propto \frac{1}{|\bR|^{2.25\pm0.05}} \text{.}
\end{equation}

\begin{figure}
\includegraphics[width=3in]{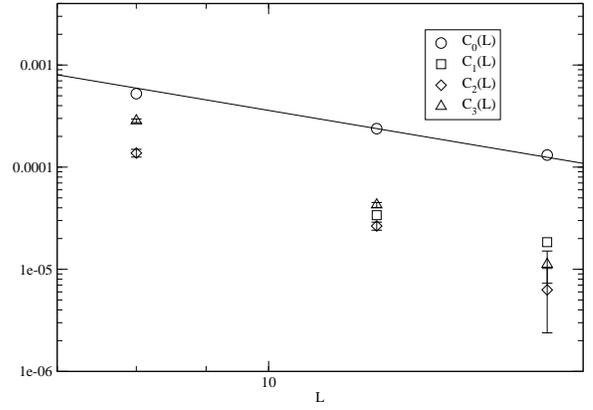}
\caption{Plot of the coefficients $c_i(L)$, for $L = 8, 12, 16$.  The value of $c_1(8)$ is not shown, as it is negative.  Error bars are on the order of the symbol size or smaller, except where shown.  The straight line is a plot of the function $e^{-2.75}/L^{2.25}$.}
\label{fig:bcoplot}
\end{figure}

Surprisingly, this power-law decay does \emph{not} seem to correspond to the Hastings VBS state.  To better understand the long-distance decay of the correlations, we have computed ${\cal C}^1(\bR)$ for a few values of $\bR$ near $(L/2) (\ba_1 + \ba_2)$ -- note that, by reflection symmetry, ${\cal C}^1(\frac{L}{2}, \frac{L}{2}) = {\cal C}^1(0, \frac{L}{2})$.  We assume the long-distance correlations are dominated by the three M-points and the $\Gamma$-point, so that, near $\bR = (L/2)(\ba_1 + \ba_2)$,
\begin{equation}
{\cal C}^1(\bR) \approx \sum_{i = 0}^3 c_i(L) e^{i \bQ_i \cdot \bR} \text{,}
\end{equation}
where $\bQ_0 = 0$, and $\bQ_i$ for $i = 1,2,3$ are the three M-point wavevectors.  The coefficients $c_i(L)$ can then be obtained from the four values ${\cal C}^1(\frac{L}{2}, \frac{L}{2})$ , ${\cal C}^1( \frac{L}{2}+ 1,  \frac{L}{2})$, ${\cal C}^1( \frac{L}{2},  \frac{L}{2} + 1)$
and ${\cal C}^1( \frac{L}{2} -1,  \frac{L}{2}+1)$, and are plotted in Fig.~\ref{fig:bcoplot}.  The coefficient $c_0(L)$ is significantly larger than the M-point coefficients, and behaves consistently with the same power-law decay found for ${\cal C}^i(\bR)$ in Fig.~\ref{bond_edge_plot}.

The fact that the dominant bond correlations lie near $\bq = 0$, rather than at the M-points, is puzzling.  Note that the spin singlet monopoles $w_i$, even if they are even under time reversal ($t_w = 1$), carry crystal momenta at the M-points and thus do not contribute to the $\bq = 0 $ correlations.  One possibility is that the dominant long distance bond correlations are still not evident for $L_1 = L_2 = 20$.  If the dominant long distance correlations really are at $\bq = 0$, then either there is another operator in the ASL leading to these correlations that has been missed so far, or the projected wavefunction simply does not give a good description of the ASL critical behavior.  Since, to our knowledge, virtually nothing is known analytically about the critical properties of Gutzwiller projected wavefunctions in two dimensions, it is impossible at the moment to decide among these possibilities.

\begin{table}
\begin{center}
\begin{tabular}{|r|r|r|r|r|r|}
\hline$L$ & $4$ & $8$ & $12$ & $16$ & $20$ \\ \hline
$\mathcal{C}^1(0,\frac{L}{2})$ & $59.4(4)$ & $8.0(3)$ & $2.9(1)$ & $1.51(9)$ & $0.91(9)$ \\ \hline
$\mathcal{C}^1(\frac{L}{2},0)$ &$-61.8(6)$ & $18.8(4)$ & $-7.4(1)$ & $-3.9(1)$ & $-2.6(1)$ \\ \hline
$\mathcal{C}^2(0,\frac{L}{2})$ & $16.7(9)$ & $-1.9(3)$ & $-1.5(1)$ & $-1.35(7)$ & $-0.98(8)$ \\ \hline
$\mathcal{C}^2(\frac{L}{2},0)$ & $83.2(8)$ & $19.6(2)$ & $7.69(7)$ & $4.11(7)$ & $2.39(6)$ \\ \hline
$\mathcal{C}^3(0,\frac{L}{2})$ & $36.8(5)$ & $9.9(2)$ & $3.83(6)$ & $1.99(6)$ & $1.23(6)$ \\ \hline
\end{tabular}
\end{center}
\caption{The bond-bond correlation in units of $10^{-4}$ at half of the system size.}\label{bond_edge_data}
\end{table}

\section{Discussion}
\label{sec:discussion}

Here, we discuss our results in the context of experiments on herbertsmithite,  then conclude with a discussion of open theoretical issues.

\begin{figure}
 \includegraphics[width=0.4\textwidth]{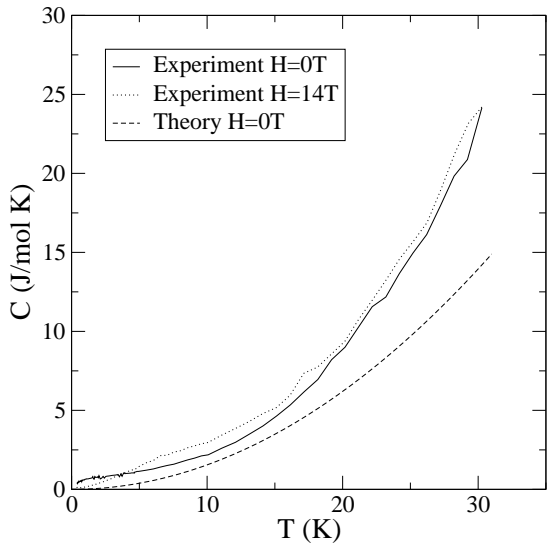}
\caption{Herbertsmithite specific heat in zero and $14$ Tesla magnetic field (courtesy of J. Helton and Y.S. Lee\cite{helton07}) together with the mean-field theoretical curve. Note that the theoretical curve is not a fit to the experimental data -- there is no tunable parameter.  We have assumed $J / k_B = 200 {\rm K}$ and used the estimate of $v_F$ (equivalently, of the hopping parameter $\chi$) in Sec.~\ref{sec:fermi-velocity}. The specific heat is per mole of formula units.}
\label{fig:specific-heat}
\end{figure}

The estimate of $v_F$ provided by the projected wavefunction calculations of Sec.~\ref{sec:fermi-velocity} allows us to calculate the specific heat $C(T)$ and magnetic susceptibility $\chi(T)$ in mean-field theory, and compare to the experimental data.  The results are
\begin{eqnarray}
\chi(T) &=& \frac{3.2 \mu_B^2}{J^2} (k_B T) \\
C(T) &=& \frac{74.6 N_A k_B^3}{J^2} T^2 \text{.}
\end{eqnarray}
Here, $\chi(T)$ is the susceptibility per Cu site, and $C(T)$ is the specific heat per mole of formula units.
The only free parameter is the exchange energy $J / k_B$, which we take to be $200$K.  We note that the $1/N$ corrections to the coefficients of $C(T)$ and $\chi(T)$ have been calculated in Ref.~\onlinecite{kaul08}.  Because of the crude nature of our estimate for $v_F$, we have not made use of those results in the above formulae.  $C(T)$ is plotted in Fig.\ref{fig:specific-heat} and agrees well with the data for $T \lesssim 30 {\rm K}$.  

For the susceptibility, we take the point of view that it is dominated by impurities at low temperature.  A better estimate of the intrinsic susceptibility of the kagome Heisenberg model is given by the NMR Knight shift.  In both Refs.~\onlinecite{imai08} and~\onlinecite{olariu08}, a component of NMR spectrum was found to have a shift that decreased at low temperature, and was argued to correspond more closely to the intrinsic susceptibility.  In the case of Ref.~\onlinecite{olariu08}, saturation is observed at the lowest temperatures.  In both cases, the slope of the decrease is consistent with that of the  calculated $\chi(T)$, within a factor of two.  In order to make this comparison, one must convert Knight shift to units of magnetic susceptibility -- this can be done using the quoted hyperfine constants in Refs.~\onlinecite{imai08} and~\onlinecite{olariu08}.  The saturation observed in Ref.~\onlinecite{olariu08} may be due to DM interaction or coupling of impurities to the bulk kagome layers.

We note that the physics of the ASL is expected to be relevant at temperatures below the spinon bandwidth $W \lesssim J$, which we estimate to be $W \approx 0.25 J$ following the analysis of Sec.~\ref{sec:fermi-velocity} (we define $W$ to be the difference in energy from the Dirac points to the top of the lowest empty band).  Taking $J / k_B \approx 200 {\rm K}$, the temperature $W/k_B \approx 50 {\rm K}$ sets the correct scale for the onset of the downturn in the Knight shift.

Perhaps the most striking prediction is the presence of gapless spin-triplet excitations at the $\Gamma$ and M points in the Brillouin zone.  If sufficiently large single crystals become available, this could be tested with inelastic neutron scattering.  Such neutron experiments would also allow detection of the magnetic competing orders at both the $\Gamma$ and M points; a strong test for the presence of the ASL would be a verification of the scaling forms for $\chi''(\bq, \omega)$ given in Sec.~\ref{sec:detection}.  The NMR relaxation rate also provides, in principle, a probe of the magnetic competing orders of the ASL via the power law behavior $1/T_1 \propto T^{\eta}$, where $\eta = 2 \Delta_N - 1$, assuming $\Delta_N < \Delta_m$ (see Sec.~\ref{sec:detection}).    Such power-law behavior has been observed\cite{imai08,olariu08}, with $\eta \approx 0.5$ (which corresponds to $\Delta_N \approx 0.75$).  However, the power law seems to depend on magnetic field,\cite{imai08} suggesting that the observed relaxation rate is dominated by magnetic impurities.

As discussed in Sec.~\ref{sec:detection}, the VBS competing order may be probed via inelastic X-ray or neutron scattering measurements of the lineshape of an appropriate optical phonon.  Again, this would require single crystal samples to be prepared.

Since the spinons carry entropy, we expect they will contribute to the thermal conductivity $\kappa$. Measurement of $\kappa$ will be important in herbertsmithite; because it distinguishes between localized and delocalized gapless excitations, it should be particularly helpful in understanding the observed spin liquid behavior.  It will also be important to develop a theoretical understanding of thermal conductivity in the ASL.

All our predictions for herbertsmithite are tempered by the twin complications of impurities and DM interaction.  To the extent that magnetic impurities play a dominant role, their effect may be reduced by applying a large enough magnetic field to polarize them.  For example, measurements of specific heat\cite{helton07} and susceptibility\cite{bert07} suggest that magnetic impurities are polarized for a field of $H = 14\, {\rm T}$ at temperature $T \approx 2\, {\rm K}$.  Assuming that the coupling of magnetic impurities to the bulk kagome spins is characterized by an energy $J_{{\rm imp}} \ll J$, then, for temperature in the range $J_{{\rm imp}} \ll T \ll W$, the effect of impurities on the intrinsic physics of the kagome layers should not be severe.  However, in this regime one needs to focus on probes where impurity and intrinsic contributions can be separated.

An understanding of the effects of DM interaction would be greatly aided by a better knowledge of its magnitude.  To this end, it will be important to measure the anisotropy of $\chi(T)$ when single crystal samples become available.  If it is possible to have a reasonable range of temperature where $D_p, D_z \ll T \ll W$, then DM interaction should not be important within this range.  For $k_B T \lesssim D_p, D_z$, we argued in Sec.~\ref{sec:dm-interaction} that DM interaction will induce spontaneous breaking of time reversal symmetry.  It would be interesting to look for this at low temperature in herbertsmithite.  We also note that, based on the local structure of the Cu-O-Cu bonds in herbertsmithite, we expect 
$D_z < D_p$, which is consistent with the results of Refs.~\onlinecite{rigol07a} and~\onlinecite{rigol07b}.  It may be the case that $D_z$ is small enough that it can be ignored.

We conclude by mentioning some of the open theoretical issues relevant to the present study.  It is important to develop a better understanding of the effects of impurities on the ASL.  The physics of single, nonmagnetic impurities have been studied in Ref.~\onlinecite{kolezhuk06}.  What is still needed is a treatment of magnetic impurities, and an understanding beyond the single impurity level.  It would also be useful to make a systematic study of slave fermion mean-field states including DM interaction.

It would be helpful to understand the critical properties of the ASL better for $N_f = 4$.  For example, it might be possible to calculate the exponent $\eta_N$ by numerical simulations of the effective field theory.  In this paper, we have used calculations in the projected wavefunction approach to try to estimate critical exponents -- a better understanding of the criticality in projected wavefunctions, and its relation to that of the ASL, would also be helpful.

\acknowledgments{We thank J. Helton, T. Imai, Y. S. Lee, C. Lhuillier, G. Misguich, M. Rigol, T. Senthil and A. Vishwanath for discussions.  This research is supported by NSF Grant Nos. DMR-0517222 (P. A. L.) and DMR-0706078 (X.-G. W.).}

\appendix
\section{Continuum limit of the mean-field state}
\label{app:cont-limit}

In this appendix, we first solve the mean-field Hamiltonian Eq.~(\ref{eqn:mft-hamiltonian}), and discuss its band structure.  Then, focusing on the low-energy excitations near the Dirac nodes, we take the continuum limit and demonstrate explicitly the relationship between the continuum and lattice spinon fields.  The realization of the microscopic symmetries (\emph{e.g.} space group) in the continuum theory depends crucially on these results; this is discussed in Appendix~\ref{app:symmetries}.

We work with the 6-site unit cell as shown in Fig.~\ref{fig:big-unitcell}.  Note that the 6-site unit cell is used in order to accommodate the background $\pi$-flux per hexagon of the ${\rm U}(1)$ gauge field; however, translation symmetry is not broken, and the true, or underlying, unit cell has 3 sites as usual.
Unit cells are labeled by Bravais lattice vectors 
$\vR = 2 n_1 \ba_1 + n_2 \ba_2$, where $n_1$ and $n_2$ are integers, and the primitive vectors are
$2 \ba_1 = 2 \bx$ and $\ba_2 = (1/2) \bx + (\sqrt{3}/2) \by$.  Note that we use the symbol $\bR$ for lattice vectors of the underlying 3-site kagome unit cell, and $\vR$ for the enlarged 6-site unit cell.
 Within each unit cell, sites are labeled as shown in Fig.~\ref{fig:big-unitcell}.  Dropping the spin index, the Hamiltonian may be written
\begin{eqnarray}
{\cal H}_{{\rm MFT}}  &=& - t \sum_{\vR} \Big\{
f^\dagger_{\vR 0} [ f^{\vphantom\dagger}_{\vR 1} + f^{\vphantom\dagger}_{\vR 2}] + f^\dagger_{\vR 1} [f^{\vphantom\dagger}_{\vR 2} - f^{\vphantom\dagger}_{\vR 3}] \nonumber \\
&+& f^\dagger_{\vR 2} [ - f^{\vphantom\dagger}_{\vR + \ba_2, 0} - f^{\vphantom\dagger}_{\vR - \ba_1 + \ba_2, 4} ] \nonumber \\
&+& f^\dagger_{\vR 3} [ f^{\vphantom\dagger}_{\vR 4} + f^{\vphantom\dagger}_{\vR 5}] 
+ f^\dagger_{\vR 4} [ f^{\vphantom\dagger}_{\vR 5} + f^{\vphantom\dagger}_{\vR + \ba_1, 0} ]  \nonumber \\
&+& f^\dagger_{\vR 5} [ f^{\vphantom\dagger}_{\vR + \ba_2, 3} - f^{\vphantom\dagger}_{\vR + \ba_2, 1} ] + \text{H.c.} \Big\}
\end{eqnarray}
We shall always take $t > 0$; in fact, this involves no loss of generality (see the discussion at the end of this appendix).

Defining the Fourier transform by
\begin{equation}
f_{\vR i} = \frac{1}{\sqrt{N_c}} \sum_{\bk} e^{i \bk \cdot \vR} f_{\bk i} \text{,}
\end{equation}
where $N_c$ is the number of unit cells, we may go to momentum space and write the Hamiltonian as
${\cal H}_{{\rm MFT}} = t \sum_{\bk} f^\dagger_{\bk i} H(\bk)_{i j} f^{\vphantom\dagger}_{\bk j}$, where
\begin{widetext}
\begin{equation}
- H(\bk) = \left( \begin{array}{cccccc}
0 & 1 & (1 - K_2^*) & 0 & K_1^* & 0 \\
1 & 0 & 1 & -1 & 0 & -K_2^* \\
(1 - K_2) & 1 & 0 & 0 & - K_1^* K_2 & 0 \\
0 & -1 & 0 & 0 & 1 & (1 + K_2^*) \\
K_1 & 0 & - K_1 K_2^* & 1 & 0 & 1 \\
0 & -K_2 & 0 & (1 + K_2) & 1 & 0
\end{array} \right) \text{.}
\end{equation}
\end{widetext}
Here, we have defined $K_1 = e^{2 i \bk \cdot \ba_1}$ and $K_2 = e^{i \bk \cdot \ba_2}$.

The primitive vectors of the reciprocal lattice are chosen to be $\boldsymbol{b}_1/2 = \pi (1, - 1/ \sqrt{3})$ and $\boldsymbol{b}_2 = 2\pi(0, 2\pi/ \sqrt{3})$.  (Again, the reciprocal lattice primitive vectors for the underlying unit cell are $\boldsymbol{b}_1$ and $\boldsymbol{b}_2$.)  The Brillouin zone can be chosen as shown in Fig.~\ref{fig:unitcell-and-bz}.  The two highest energy bands of $H(\bk)$ are completely flat with energy $2 t$.  The lowest empty and highest filled bands meet at Fermi points located at $\pm \bQ$; the Fermi energy is $\epsilon_F = t(\sqrt{3} - 1)$.

We focus on the low-energy excitations near the Dirac nodes, and so confine our attention to the two bands that touch at $\epsilon_F$.  We choose the following basis for eigenvectors of $H(\pm \bQ)$ at the Fermi energy,
\begin{widetext}
\begin{eqnarray}
(e^{+}_1)^T &=& \frac{1}{\sqrt{6}} \left( \begin{array}{cccccc}
- e^{- i \pi /24} & \sqrt{2} e^{-11\pi i / 24} & e^{i \pi /8} & e^{-i \pi / 24} & 0 & e^{-3 \pi i /8}
\end{array} \right) \\
(e^{+}_2)^T &=& \frac{1}{\sqrt{6}} \left( \begin{array}{cccccc}
 - e^{i \pi / 24} & 0 & e^{-5 \pi i /8} & - e^{i \pi / 24} & \sqrt{2} e^{11\pi i / 24} & e^{- i \pi / 8} 
\end{array} \right) \\
(e^{-}_1)^T &=& \frac{1}{\sqrt{6}} \left( \begin{array}{cccccc}
 -e^{- i \pi / 24} & 0 & e^{5 \pi i/ 8} & -e^{-i \pi / 24} & \sqrt{2} e^{-11\pi i /24} & e^{i \pi / 8} 
\end{array} \right) \\
(e^{-}_2)^T &=& \frac{1}{\sqrt{6}} \left( \begin{array}{cccccc}
 e^{i \pi / 24} & \sqrt{2}e^{-13 \pi i / 24} & - e^{- i \pi / 8} & - e^{i \pi /24} & 0 & e^{-5\pi i /8}
\end{array} \right) \text{,}
\end{eqnarray}
\end{widetext}
so that $H(\pm \bQ) e^{\pm}_i = \epsilon_F e^{\pm}_i$.  We want to write an effective Hamiltonian for these states, for small deviations of the momentum from the nodal points.  This can be done using first order perturbation theory, which leads us to write down the effective Hamiltonian
\begin{equation}
[H^{\pm}]_{i j} = (e^{\pm}_i)^\dagger D_{\pm}(\bq) e^{\pm}_{j} \text{,}
\end{equation}
Here, $D_{\pm}(\bq)$ is given by $H(\pm \bQ + \bq) - H(\pm \bQ)$, keeping only terms first order in $\bq$.  The result is
\begin{equation}
H^{\pm}(\bq) = \frac{1}{\sqrt{2}} \big[ q_x \tau^1  + q_y \tau^2 \big] \text{,}
\end{equation}
which is nothing but the Hamiltonian for massless Dirac fermions in two spatial dimensions.  Note that the velocity of the Dirac fermions is isotropic, \emph{i.e.} it does not depend on direction in $k$-space.

We can use these results to define continuum fermion fields.  Restoring the spin index $\alpha$, we write
\begin{equation}
\psi_{\alpha, \pm}(\bq) \sim \left( \begin{array}{c}
(e^{\pm}_1)^*_i f_{\pm \bQ + \bq, i, \alpha} \\
(e^{\pm}_2)^*_i f_{\pm \bQ + \bq, i, \alpha} 
\end{array} \right) \text{.}
\end{equation}
These fields obey the continuum, second-quantized Dirac Hamiltonian
\begin{equation}
{\cal H}_{{\rm Dirac}} = v_F \int \frac{d^2 \bq}{(2\pi)^2} \, \psi^\dagger_{\alpha a} \big( q_x \tau^1  + q_y \tau^2 \big) \psi^{\vphantom\dagger}_{\alpha a} \text{.}
\end{equation}

We remark that the sign of the spinon hopping $t$ is unimportant for all of our results.  From the point of view of the projected wavefunction, this can be observed by noting that $t \to -t$ under a particle-hole transformation of the lattice spinons, $f_{\br \alpha} \to (i \sigma^2)_{\alpha \beta} f^\dagger_{\br \beta}$.  Because this is an ${\rm SU}(2)$ gauge transformation, it leaves the wavefunction invariant.  To understand this from the effective field theory point of view,  we consider the lattice gauge theory Hamiltonian of Eq.~(\ref{eqn:effective-lattice-hamiltonian}), and explicitly keep track of its dependence on $t$ by writing
${\cal H}_{{\rm eff}} = {\cal H}_{{\rm eff}}(t)$.  Noting that both the electric field and vector potential are odd under particle-hole transformation, we have ${\cal H}_{{\rm eff}}(t) \to {\cal H}_{{\rm eff}}(-t)$.  Let ${\cal O}$ be any combination of spin operators (it need not be a local operator in space or time).  ${\cal O}$ is invariant under ${\rm SU}(2)$ gauge transformations, and, in particular, is invariant under particle-hole transformation.  Therefore the expectation value of ${\cal O}$ will be identical if calculated using either ${\cal H}_{{\rm eff}}(t)$ or ${\cal H}_{{\rm eff}}(-t)$, and we conclude there is no physical distinction between these two effective Hamiltonians.   For simplicity, therefore, we always choose $t > 0$ as we have done above.

\section{Symmetries}
\label{app:symmetries}

Here, we outline the procedure for calculating the realization of microscopic symmetries for the continuum Dirac field $\Psi$.  The basic idea is to diagonalize the mean-field Hamiltonian for a finite system, and extract the needed information from properties of the nodal wavefunctions.  The necessary manipulations are easily carried out with standard symbolic or numerical packages for linear algebra computations.  

We work with the 6-site unit cell, and consider a system with periodic boundary conditions in the $2 \ba_1$ and $\ba_2$ directions, so that $\vR$, $\vR + 2 L_1 \ba_1$ and $\vR + L_2 \ba_2$ are all identified.  We label sites by $\br$ or, equivalently, by the pair $(\vR, i)$.  The number of sites is then $N_s = 6 N_c = 6 L_1 L_2$, and the Hamiltonian is a $N_s \times N_s$ matrix defined by $(H)_{\br \br'} = \mp 1$ for $\br$ and $\br'$ nearest neighbors, and zero otherwise.  The negative (positive) sign is taken for the thick (thin) bonds in Fig.~\ref{fig:big-unitcell}.  $L_1$ and $L_2$ must be chosen so that the nodal wavevectors $\pm \bQ$ are in fact present in the Brillouin zone of the finite size system.

The spin plays no role in these manipulations, so we drop the spin index, and we can then think of the continuum Dirac field as a four-component object.  We write
\begin{equation}
\Psi_{a p} (\bq = 0) = \sum_{\br} \Phi^*_{a p} (\br) f_{\br} \text{,}
\end{equation}
where $a = +, -$ is the nodal index, $p = 1,2$ is the index in the two-component Dirac space.  The $\Phi_a$ are the nodal wavefunctions, satisfying $H \Phi_a = \epsilon_F \Phi_a$, and are given by
\begin{equation}
\Phi_{a p} (\vR , i) = \frac{e^{i a \bQ \cdot \vR} e^{a}_p (i)}{\sqrt{N_c}} \text{.}
\end{equation}

Now consider a symmetry operation $S$, with the following action on the fermion fields:
\begin{equation}
S : f_{\br} \to \pi_{\br} f_{S(\br)} \text{,}
\end{equation}
where in the present case we can take the gauge transformation $\pi_{\br} = \pm 1$.  This induces the following action on the wavefunctions:
\begin{equation}
(S \Phi_a)\big( S(\br) \big) = \pi_{\br} \Phi_a(\br) \text{.}
\end{equation}
This allows us to define the matrix of the symmetry operation by
\begin{equation}
(S)_{ S(\br), \br } = \pi_{\br} 
\end{equation}
for all $\br$, with all other elements zero.

Next, we can express the action of the symmetry on the nodal wavefunctions as
\begin{equation}
S \Phi_a = c_{a b} \Phi_b \text{,}
\end{equation}
where the coefficients $c_{a b}$ can be explicitly computed by taking inner products.  Translating this into the action on the fermion field, we have:
\begin{equation}
S : \Psi_a \to c^*_{a b} \Psi_b \text{,}
\end{equation}
which gives us the desired result.

\section{Group theory of the kagome lattice}
\label{app:grouptheory}

\subsection{Outline}

Here, we work out some details of the group theory and representation theory of the kagome lattice space group.  The goal is to understand the action of the space group on objects invariant under translations by $2 \ba_1$ and $2 \ba_2$, which is true of the competing orders within the ASL, and to that end we define a ``reduced'' space group.  We use these results to classify all possible site and bond ordering patterns invariant under $T_{2 \ba_1}$ and $T_{2 \ba_2}$.

\subsection{Point group}

\begin{figure}
\includegraphics[width=2in]{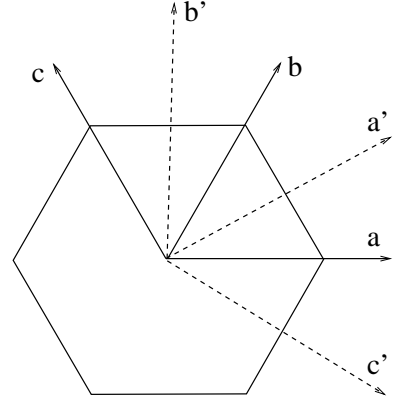}
\caption{2-fold rotation axes for the symmetries of the hexagon.  Equivalently, these can be thought of as mirror symmetries in the plane.}
\label{fig:hexagon-symms}
\end{figure}

The point group of the kagome lattice is $D_6$, which is the symmetry group of a regular hexagon.  The group $D_6$ has 12 elements.  We define the elements of $D_6$ by operations on the hexagon shown in Fig.~\ref{fig:hexagon-symms}.  $D_6$ is generated by $C_6 = R_{\pi/3}$, which is a counterclockwise rotation by $\pi/3$ about the $z$-axis piercing the center of the hexagon, and $C_a = {\cal R}_y$, which is a $\pi$ rotation about the $a$-axis as shown in Fig.~\ref{fig:hexagon-symms} (this is equivalent to a mirror symmetry in the plane).  The group is completely specified by the relations
$C_6^6 = C_a^2 = (C_6 C_a)^2 = 1$.\cite{hamermesh62}

The 12 elements of $D_6$ are
\begin{equation}
D_6 = \{ 1, C_6, C_6^2, C_6^3, C_6^4, C_6^5, C_a, C_{a'}, C_b, C_{b'}, C_c, C_{c'} \} \text{.}
\end{equation}
Here, $C_{a'}$, $C_{b}$, etc, are rotations by $\pi$ about the axes shown in Fig.~\ref{fig:hexagon-symms}.  
$D_6$ has the 6 conjugacy classes
${\cal C}_1 = \{ 1 \}$,  
${\cal C}_6 = \{ C_6, C_6^5 \}$,
${\cal C}^2_6 = \{ C^2_6, C^4_6 \}$,
${\cal C}^3_6 = \{ C^3_6 \}$,
${\cal C}_a = \{ C_a, C_b, C_c \}$, and
${\cal C}_{a'} = \{ C_{a'}, C_{b'}, C_{c'} \}$.
The character table is given in Table~\ref{tab:d6-character}.\cite{hamermesh62}  The faithful representation $E_1$ is obtained by straightforwardly representing $D_6$ in terms of $2 \times 2$ rotation and reflection matrices.  $A_2$ can be obtained from this representation by taking the determinant -- it distinguishes rotations from mirror symmetries in the plane.  Note that $E_2 = B_1 \otimes E_1 = B_2 \otimes E_1$.  

\begin{table}
\begin{tabular}{l | c | c | c  | c | c | c |}
Rep. & ${\cal C}_1$ & ${\cal C}^3_6$ & ${\cal C}^2_6$ & ${\cal C}_6$ & ${\cal C}_a$ & ${\cal C}_{a'}$ \\
\hline
$A_1$ & 1 & 1 & 1 & 1 & 1& 1 \\
\hline
$A_2$ & 1 & 1 & 1 & 1 & -1& -1 \\
\hline
$B_1$ & 1 & -1 & 1 & -1 & 1& -1 \\
\hline
$B_2$ & 1 & -1 & 1 & -1 & -1& 1 \\
\hline
$E_1$ & 2 & -2 & -1 & 1 & 0 & 0 \\
\hline
$E_2$ & 2 & 2 & -1 & -1 & 0 & 0 \\
\hline
\end{tabular}
\caption{Character table for $D_6$.\cite{hamermesh62}  The first column labels the representation, and the following columns give the values of the character on each conjugacy class.}
\label{tab:d6-character}
\end{table}

\subsection{Space group}

The space group of the kagome lattice is generated by translations, and the $D_6$ point group operations preserving the center of a particular hexagon.  A general element of the space group can by represented by the Seitz operator $\{ R | \boldsymbol{t} \}$, where $R \in {\rm O}(2)$, and $\boldsymbol{t} = n_1 \ba_1 + n_2 \ba_2$ is a lattice translation vector.  This operator is defined in terms of its action on a lattice point $\br$:
\begin{equation}
\{ R | \boldsymbol{t} \} \br = R \br + \boldsymbol{t} \text{.}
\end{equation}
The entire space group can be generated by $\{ C_6 | 0 \} = R_{\pi/3}$, $\{ C_a | 0 \} = {\cal R}_y$ and $\{ 1 | \ba_1 \} = T_{\ba_1}$.

We now recall some general facts about space groups.  Denote the space group by $G_s$.  The translation group $G_t$ is a normal subgroup (\emph{i.e.} $g G_t g^{-1} = G_t$ for all $g \in G_s$).  The factor group $G_p = G_s / G_t$ is the point group.

\subsection{Reduced space group}

Suppose we are interested in the transformation properties of some object invariant under translation by a certain number of lattice vectors, say by $m_1 \ba_1$ and $m_2 \ba_2$.  So, acting on this object, we have $T^{m_1}_{\ba_1} = T^{m_2}_{\ba_2} = 1$.  Let us suppose $m_1 = m_2 = m$, and define the group of translations leaving our object invariant by
\begin{equation}
G_{tm} = \Big\{ \{ 1 | m (n_1 \ba_1 + n_2 \ba_2) \} ; n_1, n_2 \in {\mathbb Z} \Big\} \text{.}
\end{equation}
Because this set of vectors is invariant under the kagome space group, $G_{tm}$ forms a normal subgroup of $G_s$.  (Note this would not be the case if $m_1 \neq m_2$.)  Therefore we can define the reduced space group as the factor group $G_{sm} = G_s / G_{tm}$.  The reduced space group completely describes the action of the space group on our object of interest.

The enhanced fermion bilinears in our kagome algebraic spin liquid are invariant under translation by $2 \ba_1$ and $2 \ba_2$.  So, to understand their transformations under the space group, we consider $m=2$ and $G_{s2}$.  $G_{s2}$ can be interpreted as the kagome space group, acting on objects with crystal momentum lying at the $\Gamma$ point ($\bq = 0$), or one of the three M-points.  Alternatively, $G_{s2}$ gives the action of the space group on any ordering pattern on the kagome lattice with the 12-site $2 \times 2$ unit cell (\emph{i.e.} 2 unit cells by 2 unit cells) invariant under $T_{2 \ba_1}$ and $T_{2 \ba_2}$.

We shall work out some properties of $G_{s2}$ and determine its character table.  Elements of $G_{s2}$ can be represented again by Seitz operators $\{ R | \boldsymbol{t} \}$, where now we restrict
$\boldsymbol{t} = 0, \ba_1, \ba_2, \ba_1 + \ba_2$.  The Seitz operators multiply by the usual rules, except that if one obtains a vector $\boldsymbol{t}$ violating the restriction above, we add and subtract integer multiples of $2 \ba_1$ and $2 \ba_2$ so that the restriction is satisfied -- this is just the usual way to multiply elements of a factor group.  $G_{s2}$ has 48 elements.  There is a translation subgroup of $G_{s2}$, which we denote by $T_2 = \big\{ \{ 1 | \boldsymbol{t} \} , \boldsymbol{t} = 0, \ba_1, \ba_2, \ba_1 + \ba_2 \big\}$.  $T_2$ is a normal subgroup, and clearly $T_2 \simeq Z_2 \times Z_2$.  The factor group $G_{s2} / T_2$ is just the point group $D_6$ again.  $G_{s2}$ has 10 conjugacy classes; each class is listed in Table~\ref{tab:gs2-conj-classes}, together with its size and a representative element.

\begin{table}
\begin{tabular}{l | c | c | c | c | c |}
Conj. Class & ${\cal C}_1$ & ${\cal C}_t$ & ${\cal C}^3_6$ & ${\cal C}^3_{6t}$ & ${\cal C}^2_6$ \\
\hline
Number elts. & 1 & 3 & 1 & 3 & 8  \\
\hline
Representative & 1 & $\{ 1 | \ba_1 \}$ & $\{ C^3_6 | 0 \}$ & $\{ C^3_6 | \ba_1 \}$ &
$\{ C^2_6 | 0 \}$    \\
\hline
& & & & &
\\
\hline
Conj. Class & ${\cal C}_6$ & ${\cal C}_a$ & $\tilde{C}_a$ & ${\cal C}_{a'}$ & $\tilde{C}_{a'}$ \\
\hline
Number elts. & 8 & 6 & 6 & 6 & 6 \\
\hline
Representative & $\{ C_6 | 0 \}$ & $\{ C_a | 0 \}$ & $\{ C_a | \ba_2 \}$  & $\{ C_{a'} | 0 \}$ &
$\{ C_{a'} | \ba_1 \}$ \\
\hline
\end{tabular}
\caption{Conjugacy classes of $G_{s2}$, shown with their sizes and a representative element.}
\label{tab:gs2-conj-classes}
\end{table}

The character table for $G_{s 2}$ is given in Table~\ref{tab:gs2-characters}.  The first six representations are obtained from those of $D_6$, exploiting the fact that $D_6 \simeq G_{s 2} / T_2$.  To work out the properties of the remaining four representations, first note that the sum of the squares of their dimensions must add up to 36.  The only possibilities of dimensions consistent with this are $\{ 3,3,3,3\}$ and $\{5,3,1,1\}$.  Suppose the second possibility occurs, and $F_1$ is a 1-dimensional representation.  Now it must be the case that $U^{F_1}(t) = s$ for all translations $t \in T_2$, where $s = \pm 1$.  This is because one-dimensional representations must be constant on conjugacy classes, and because the translations all satisfy $t^2 = 1$.  Now we cannot have $s = 1$, because then we would have obtained a distinct irreducible representation of $D_6$.  Therefore $s = -1$.  However, we have
\begin{equation}
-1 = U^{F_1}(T_{\ba_1 + \ba_2}) = U^{F_1}(T_{\ba_1}) U^{F_1}(T_{\ba_2}) = 1 \text{,}
\end{equation}
a contradiction.  So it must be the case that all $F_i$ are 3-dimensional.

\begin{table}
\begin{tabular}{l | c | c | c | c | c | c | c | c | c | c |}
Rep. & ${\cal C}_1$ & ${\cal C}_t$ & ${\cal C}^3_6$ & ${\cal C}^3_{6t}$ & ${\cal C}^2_6$
& ${\cal C}_6$ & ${\cal C}_a$ & $\tilde{C}_a$ & ${\cal C}_{a'}$ & $\tilde{C}_{a'}$ \\
\hline
$A_1$ & 1 & 1 &  1 &  1 & 1 & 1 & 1 & 1 & 1 & 1 \\
\hline
$A_2$ & 1 & 1 & 1 & 1 & 1 & 1 & -1 & -1 & -1 & -1 \\
\hline
$B_1$ & 1& 1 & -1 & -1 & 1 & -1 & 1 & 1 & -1 & -1 \\
\hline
$B_2$ & 1 & 1 & -1 & -1 & 1 & -1 & -1 & -1 & 1 &  1 \\
\hline
$E_1$ & 2 & 2 & -2 & -2 & -1 & 1 & 0 & 0 & 0 & 0 \\
\hline
$E_2$ & 2 & 2 & 2 & 2 & -1 &  -1 & 0 & 0 & 0 & 0 \\
\hline
$F_1$ & 3 & -1 & 3 & -1 & 0 & 0 & 1 & -1 & 1 & -1 \\
\hline
$F_2$ & 3 & -1 & 3 & -1 & 0 & 0 & -1 & 1 & -1 & 1 \\
\hline
$F_3$ & 3 & -1 & -3 & 1 & 0 & 0 & 1 & -1 & -1 & 1 \\
\hline
$F_4$ & 3 & -1 & -3 & 1 & 0 & 0 & -1 & 1 & 1 & -1 \\
\hline
\end{tabular}
\label{tab:gs2-characters}
\caption{Character table of $G_{s2}$.}
\end{table}

We can construct the representation $F_1$ explicitly -- it is made up of the $3 \times 3$ matrices $U^{F_1}_{i j}(S)$ describing the action of the space group on $\bN^i_A$ and $N^i_C$, which was introduced in Eq.~(\ref{eqn:F1-matrices}).  These matrices can be determined using the symmetry transformations of Sec.~\ref{sec:symmetries}, and the character of $F_1$ is then easily obtained.  Finally, we obtain the remaining representations by taking tensor products of $F_1$ with the $d=1$ irreducible representations.  Specifically, $F_2 = A_2 \otimes F_1$,
$F_3 = B_1 \otimes F_1$ and $F_4 = B_2 \otimes F_1$.

\begin{figure}
\includegraphics[width=3in]{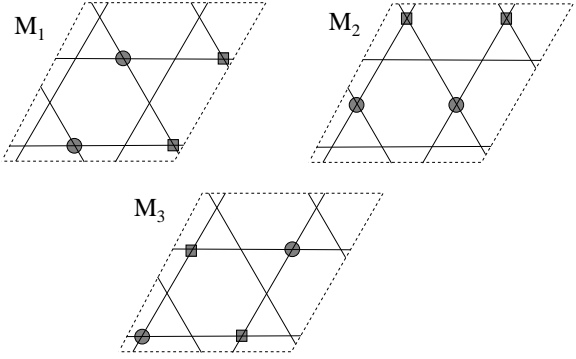}
\caption{Depiction of the three patterns of site order transforming according to the $F_1$ representation.  Each lies within the 12-site unit cell used to construct all possible site and bond ordering patterns.  Each pattern's crystal momentum lies at one of the M-points of the Brillouin zone M$_i$ (see Fig.~\ref{fig:unitcell-and-bz}).  In each case the coefficients $c_{\br}$ are zero or $\pm 1$.  The circles represent $c_{\br} = 1$, the squares $c_{\br} = -1$.}
\label{fig:f1-siteorders}
\end{figure}

\subsection{Site and bond ordering patterns}

We can use the above group-theoretic results to classify all possible ordering patterns with the 12-site unit cell shown in Fig.~\ref{fig:f1-siteorders}.  Such patterns are invariant under translations by $2 \ba_1$ or $2 \ba_2$, and thus transform under the reduced space group $G_{s2}$.  

We first focus on patterns of order that can be visualized in terms of a real field residing on the lattice sites.  We are primarily interested in collinear spin ordering patterns, and, in this case, the real field should be associated with $\langle S^z_{\br} \rangle$.
We define a 12-dimensional vector space, where basis vectors are labeled by $| \br \rangle$, and $\br$ is one of the 12 sites in the unit cell.  The action of the space group on this vector space is given by $S | \br \rangle = | S(\br) \rangle$, and we have thus constructed a (reducible) representation of the space group, which we shall call $V_s$.  The matrices for this representation can be explicitly constructed, and the character is as shown in Table~\ref{tab:VsVb-character}.  The decomposition of $V_s$ into irreducible representations is
\begin{equation}
\label{eqn:Vs-decomposition}
V_s = A_1 \oplus E_2 \oplus F_1 \oplus F_3 \oplus F_4 \text{.}
\end{equation}

An ordering pattern of the type we consider here is associated with a real linear combination of the $| \br \rangle$,
\begin{equation}
\label{eqn:site-order-vector}
| \text{site order} \rangle = \sum_{\br} c_{\br} | \br \rangle \text{,}
\end{equation}
where the coefficients $c_{\br} \sim \langle S^z_{\br} \rangle$ are the values of the ordering field.  For each irreducible representation in Eq.~(\ref{eqn:Vs-decomposition}), if we find a basis of vectors in the form Eq.~(\ref{eqn:site-order-vector}), then we have found the site ordering pattern transforming in that representation.  We are particularly interested in the site ordering pattern transforming as the $F_1$ representation, because this transforms identically to some of the competing orders in the Hastings ASL (the $\bN^i_A$ and $N^i_C$ fermion bilinears).  The patterns transforming in this representation are depicted in Fig.~\ref{fig:f1-siteorders}.

\begin{figure}
\includegraphics[width=3in]{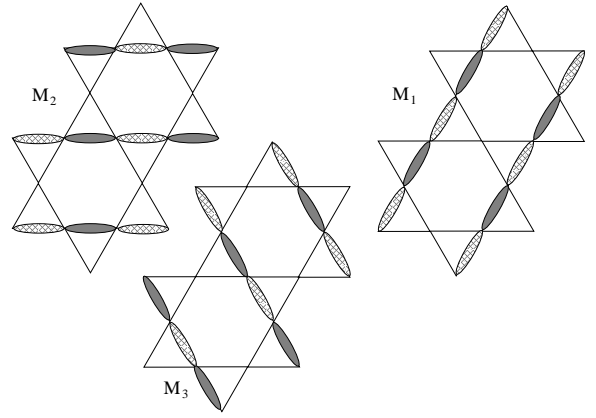}
\caption{Depiction of the three $F^1_A$ bond orders.  The patterns are labeled by their crystal momenta (Fig.~\ref{fig:unitcell-and-bz}).  The coefficients $c_b$ are zero or $\pm 1$.  Crosshatched bonds represent $c_b = 1$, and gray-shaded bonds $c_b = -1$.}
\label{fig:f1a-bondorders}
\end{figure}

\begin{table}
\begin{tabular}{l | c | c | c | c | c | c | c | c | c | c |}
Rep. & ${\cal C}_1$ & ${\cal C}_t$ & ${\cal C}^3_6$ & ${\cal C}^3_{6t}$ & ${\cal C}^2_6$
& ${\cal C}_6$ & ${\cal C}_a$ & $\tilde{C}_a$ & ${\cal C}_{a'}$ & $\tilde{C}_{a'}$ \\
\hline
$V_s$ & 12 & 0 & 0 & 4 & 0 & 0 & 2 & 0 & 2 & 0 \\
\hline
$V_b$ & 24 & 0 & 0 & 0 & 0 & 0 & 0 & 0 & 4 & 0 \\
\hline
\end{tabular}
\caption{Characters of $V_s$ and $V_b$ representations of $G_{s2}$.}
\label{tab:VsVb-character}
\end{table}

We now carry out the same analysis as above, but for ordering patterns that can be visualized in terms of a real field residing on nearest-neighbor bonds.  
There are 24 bonds in the unit cell -- to each of these we associate a vector $| b \rangle$, and bond ordering patterns correspond to real linear combinations of the form
\begin{equation}
\label{eqn:bond-order-vector}
| \text{bond order} \rangle = \sum_b c_b | b \rangle \text{.}
\end{equation}
As before, the action of the space group on this vector space is given by $S | b \rangle = | S(b) \rangle$.  We refer to the resulting 24-dimensional representation of $G_{s2}$ as $V_b$; its character is given in Table~\ref{tab:VsVb-character}.  The decomposition into irreducible representations is
\begin{equation}
V_b = A_1 \oplus B_2 \oplus E_1 \oplus E_2 \oplus F_1 \oplus F_1 \oplus F_2 \oplus F_3 \oplus F_4 \oplus F_4 \text{.}
\end{equation}
It should be noted that $F_1$ occurs \emph{twice}; this is tied to the fact that the Hastings state has three inequivalent bonds in its unit cell, as discussed below.

\begin{figure}
\includegraphics[width=3in]{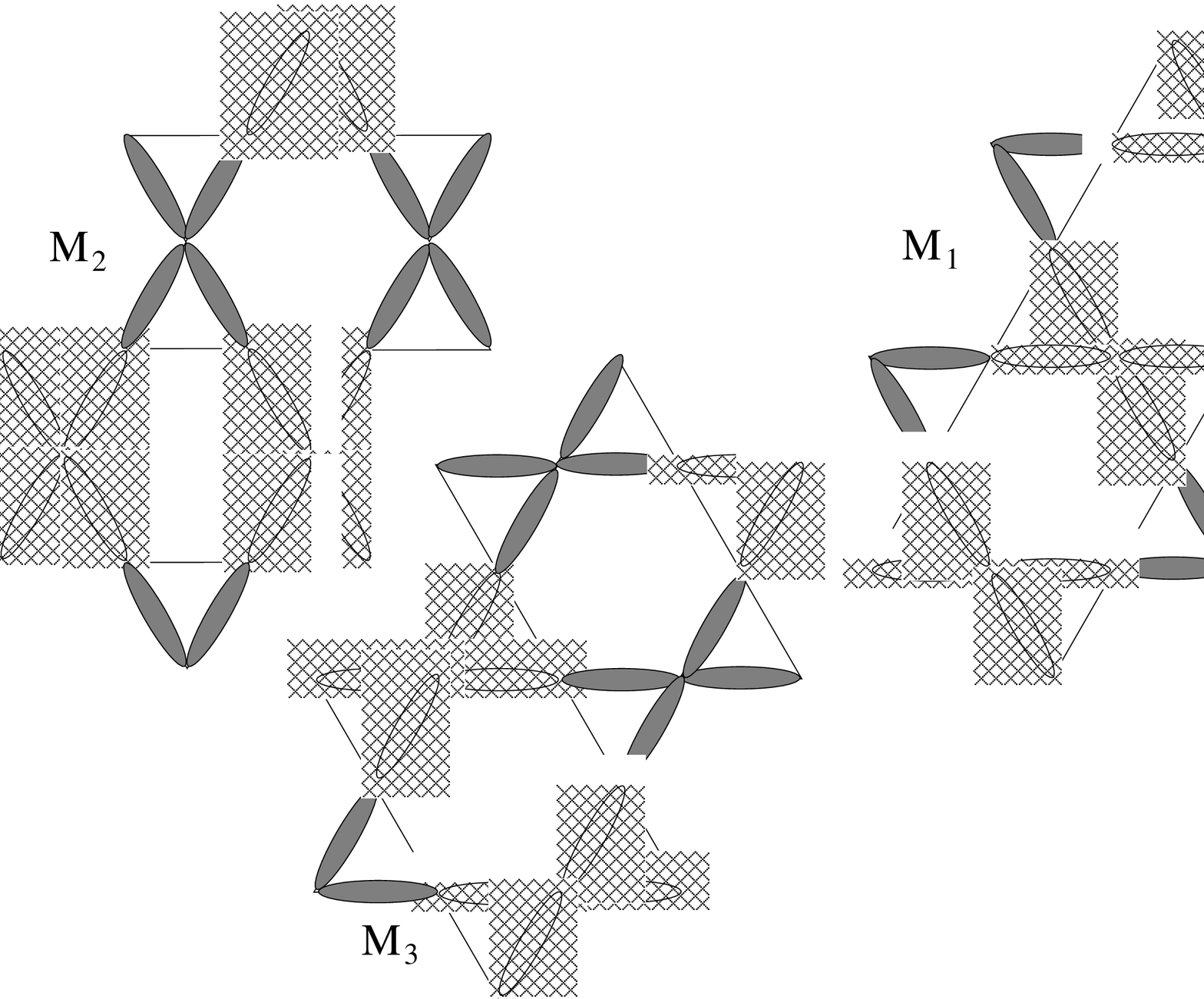}
\caption{Depiction of the three $F_1^B$ bond orders.  The patterns are labeled by their crystal momenta (Fig.~\ref{fig:unitcell-and-bz}).  The coefficients $c_b$ are zero or $\pm 1$.  Crosshatched bonds represent $c_b = 1$, and gray-shaded bonds $c_b = -1$.}
\label{fig:f1b-bondorders}
\end{figure}

\begin{figure}
\includegraphics[width=3in]{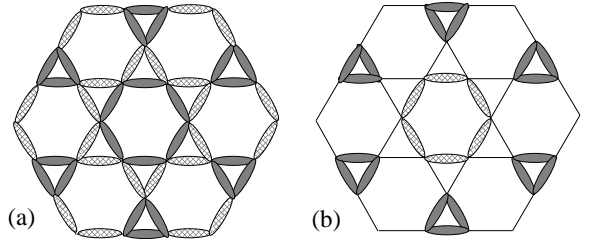}
\caption{Patterns of bond order formed by superposing the three $F^1_A$ patterns (a), and the three $F^1_B$ patterns (b).  The coefficients $c_b$ are zero or $\pm 1$.  Crosshatched bonds represent $c_b = 1$, and gray-shaded bonds $c_b = -1$.  These two patterns can in turn be superposed with the uniform state (all $c_b = 1$), to form the Hastings state, with different strengths of the three inequivalent bonds in the unit cell.}
\label{fig:bondorders-superposed}
\end{figure}

The first $F_1$ irreducible representation of bond orders ($F_1^A$) consists of the patterns shown in Fig.~\ref{fig:f1a-bondorders}, the second ($F_1^B$) of the patterns shown in Fig.~\ref{fig:f1b-bondorders}.  The $F_1^A$ bond orders can be superposed to form the pattern shown in Fig.~\ref{fig:bondorders-superposed}a, and the $F_1^B$ orders to form that shown in Fig.~\ref{fig:bondorders-superposed}b.  Formally, this superposition is achieved by taking linear combinations in the vector space defined by Eq.~(\ref{eqn:bond-order-vector}).  The Hastings state can be viewed as a superposition of the two patterns in Fig.~\ref{fig:bondorders-superposed}, together with the uniform state where all bonds have the same amplitude.  Depending on the relative weights of these three states, one constructs Hastings states with different strengths of the three inequivalent bonds in its unit cell (see Sec.~\ref{sec:fermion-bilinears}).

\section{Order parameter for $\bq = 0$ state.}
\label{app:qzero-state}

\begin{figure}
\includegraphics[width=3in]{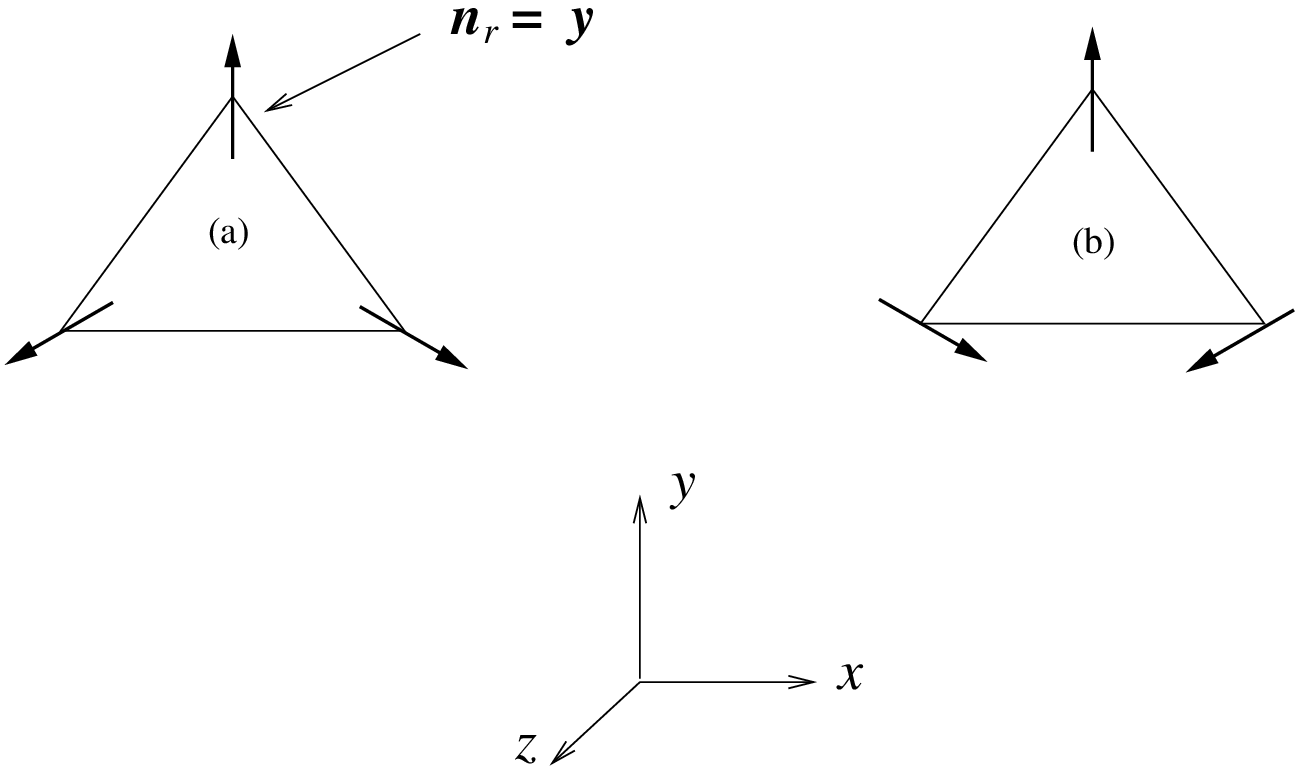}
\caption{Spin configuration on up-pointing triangles for two inequivalent $\bq = 0$ ground states.  In both triangles $\bn_r = \by$.  In triangle (a), the chirality vector $\boldsymbol{c} = \bz$, while, in triangle (b),
$\boldsymbol{c} = -\bz$.}
\label{fig:qzero-construction}
\end{figure}

Here, we construct the order parameter for the $\bq = 0$ magnetically ordered ground state of the kagome lattice, in terms of the complex vector $\bn = \bn_r + i \bn_i$.  The $\bq = 0$ state is a coplanar ordering formed by choosing the vector sum of the ordered moments on a single up-pointing triangle to be zero, and then translating this triangle to fill the rest of the lattice.  The $\bq = 0$ state is thus completely specified by the orientations of spins on a single up-pointing triangle.  We choose $\bn_r$ to be equal to the ordered moment on the ``top'' site of the up-pointing triangle, as shown in Fig.~\ref{fig:qzero-construction}.  The remaining moments are specified by the chirality vector $\boldsymbol{c}$, where $\boldsymbol{c} \cdot \bn_{r} = 0$.  Moving counterclockwise around the triangle, the spin on each site is rotated $120^{\circ}$ from the previous one about the axis given by $\boldsymbol{c}$.  This is illustrated in Fig.~\ref{fig:qzero-construction}.  We choose the imaginary part $\bn_i = \boldsymbol{c} \times \bn_r$.

The action of various symmetry operations on the order parameter $\bn$ is now easily worked out.  We have
\begin{eqnarray}
T_{\ba_1} : \bn &\to& \bn \\
R_{\pi/3} : \bn &\to& e^{2\pi i / 3} \bn \\
{\cal R}_y : \bn &\to& \bn^* \\
{\cal T} : \bn &\to& - \bn^* \text{.} 
\end{eqnarray}

\section{Fully symmetric projected wavefunctions}
\label{app:construct-prj}
In this section we discuss how to construct a Gutzwiller projected wavefunction which is fully symmetric under the space group. First, we show that a simple projected wavefunction on a finite system may break lattice symmetry. 

\begin{figure}
 \includegraphics[width=0.3\textwidth]{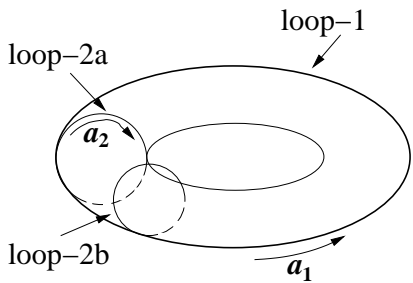}
\caption{We plot the global loops on a torus. To demonstrate the kagome lattice problem in text, we plot loop-1 which is along $\ba_1$ direction, loop-2a and loop-2b which are along $\ba_2$ direction. Loop-2a and loop-2b are separated by one unit cell spacing.}
\label{fig:loop_on_torus}
\end{figure}

Two projected wavefunctions are identical (up to a phase) if one can be transformed into another by a gauge transformation. Therefore a projected wavefunction is completely determined by the hopping magnitudes on all the bonds and the fluxes through all closed loops (assuming there is no pairing), which are both gauge invariant. On a torus the fluxes through all loops are reduced into the fluxes inside all plaquettes, and two global loops going across the two boundaries (for instance, loop-1 and loop-2a in Fig.~\ref{fig:loop_on_torus}). 

Consider the projected wavefunction for the ASL on a kagome lattice torus with $L_2$ odd; that is, an odd number of unit cells in the $\ba_2$ direction.  In Fig.~\ref{fig:loop_on_torus} we show two global loops, loop-2a and loop-2b, which are separated by one unit cell spacing.  That is, they are related by the translation $T_{\ba_1}$.) Accordingly there are $L_2$ unit cells contained in the cylinder between loop-2a and loop-2b.  Because there is $\pi$ flux through each unit cell, in total we have  $\pi L_2 = \pi \operatorname{mod} 2\pi$ flux through the cylinder. As a result, the fluxes through loop-2a and loop-2b differ by $\pi$. This means the translation symmetry along the $\ba_1$ direction is broken in the projected wavefunction.

To avoid this problem, we focus on lattices with even $L_1$ and $L_2$.
Repeating the analysis above, we conclude that translation symmetry along both $\ba_1$ and $\ba_2$ directions is preserved.  However, there are still two issues that need to be addressed.  First, we shall see that, in order to construct a symmetric wavefunction, we need to ensure that the mean-field Hamiltonian does not have fermion states lying precisely at the Dirac nodes.
Second, even if nodal states are not present, the point group symmetry may still be broken.
Both these issues are resolved by noting that one still has the freedom to choose the fluxes through the two holes of the torus. The two choices preserving time-reversal and translation symmetry are $0$ and $\pi$.  For $L_1 = L_2 = 4N$, $0$ flux corresponds to periodic boundary conditions, and $\pi$ flux to antiperiodic boundary conditions.  On the other hand, for $L_1 = L_2 = 4N + 2$, $0$ flux corresponds to \emph{antiperiodic} boundary conditions in the $\ba_1$-direction and \emph{periodic} boundary conditions in the $\ba_2$-direction.  $\pi$ flux corresponds to periodic boundary conditions in the $\ba_1$-direction, and antiperiodic boundary conditions in the $\ba_2$-direction.

If there are nodal states in the mean-field Hamiltonian, there is a ground state degeneracy arising from the different ways to fill these states, and the resulting wavefunctions transform nontrivially under microscopic symmetries.  Furthermore, we find the ground state energy is usually increased in such situations.  Therefore, we always choose the  boundary conditions to avoid the nodal fermions. We introduce the notation $[P,A]$, for example, to denote the state with periodic boundary condition in $\ba_1$ direction and anti-periodic boundary condition in $\ba_2$ direction. We find that for $L_1 = L_2 = 4 N$, $[P,A]$, $[A,P]$ and $[A,A]$ lack nodal fermions, and $[P,P]$ has nodal fermions.  For $L_1 = L_2 = (4N + 2)$, $[P,P]$, $[P,A]$ and $[A,P]$ lack nodal fermions, and $[A,A]$ has nodal fermions.

We find that in both bases, the three states avoiding nodal fermions transform into one another other under point group symmetry, thus forming a three dimensional representation of the point group. Because the $D_6$ point group only has one and two dimensional irreducible representations (see Table~\ref{tab:d6-character}), this three dimensional representation is reducible. In particular, for $L_1 = L_2 = (4 N + 2)$, the three dimensional representation is $A_2\oplus E_2$, and therefore it is impossible to use it to construct a symmetric wavefunction.  But, for $L_1 = L_2 = 4 N$, we find the representation is $A_1\oplus E_2$, and we are able to construct a symmetric wavefunction, because $A_1$ is the trivial representation. This symmetrized wavefunction, which is a linear superposition of $[P,A]$, $[A,P]$ and $[A,A]$ states, is the projected wavefunction that we used to study the spin and bond correlations in Sec.~\ref{sec:spin-correlation} and Sec.~\ref{sec:bond-correlation}.

The fact that it is impossible to construct a fully symmetric wavefunction for $L_1 = L_2 = (4 N + 2)$  from the $[P,A],[P,P], \dots$ wavefunctions does not mean it is impossible to construct a fully symmetric projected wavefunction at all.  For example, one may be able to use states with twisted boundary conditions, and obtain a linear combination invariant under both time reversal and space group symmetry. This possibility remains to be studied.  Finally, we remark that the procedure described here cannot construct fully symmetric wavefunctions on the 36-site cluster that has been extensively studied using exact diagonalization.\cite{leung93, lecheminant97, waldtmann98, sindzingre00}

\bibliography{asl-kagome}

\begin{thebibliography}{55}
\expandafter\ifx\csname natexlab\endcsname\relax\def\natexlab#1{#1}\fi
\expandafter\ifx\csname bibnamefont\endcsname\relax
  \def\bibnamefont#1{#1}\fi
\expandafter\ifx\csname bibfnamefont\endcsname\relax
  \def\bibfnamefont#1{#1}\fi
\expandafter\ifx\csname citenamefont\endcsname\relax
  \def\citenamefont#1{#1}\fi
\expandafter\ifx\csname url\endcsname\relax
  \def\url#1{\texttt{#1}}\fi
\expandafter\ifx\csname urlprefix\endcsname\relax\def\urlprefix{URL }\fi
\providecommand{\bibinfo}[2]{#2}
\providecommand{\eprint}[2][]{\url{#2}}

\bibitem[{\citenamefont{Helton et~al.}(2007)\citenamefont{Helton, Matan,
  Shores, Nytko, Bartlett, Yoshida, Takano, Suslov, Qiu, Chung
  et~al.}}]{helton07}
\bibinfo{author}{\bibfnamefont{J.~S.} \bibnamefont{Helton}},
  \bibinfo{author}{\bibfnamefont{K.}~\bibnamefont{Matan}},
  \bibinfo{author}{\bibfnamefont{M.~P.} \bibnamefont{Shores}},
  \bibinfo{author}{\bibfnamefont{E.~A.} \bibnamefont{Nytko}},
  \bibinfo{author}{\bibfnamefont{B.~M.} \bibnamefont{Bartlett}},
  \bibinfo{author}{\bibfnamefont{Y.}~\bibnamefont{Yoshida}},
  \bibinfo{author}{\bibfnamefont{Y.}~\bibnamefont{Takano}},
  \bibinfo{author}{\bibfnamefont{A.}~\bibnamefont{Suslov}},
  \bibinfo{author}{\bibfnamefont{Y.}~\bibnamefont{Qiu}},
  \bibinfo{author}{\bibfnamefont{J.-H.} \bibnamefont{Chung}},
  \bibnamefont{et~al.}, \bibinfo{journal}{Phys. Rev. Lett.}
  \textbf{\bibinfo{volume}{98}}, \bibinfo{pages}{107204}
  (\bibinfo{year}{2007}).

\bibitem[{\citenamefont{Mendels et~al.}(2007)\citenamefont{Mendels, Bert,
  de~Vries, Olariu, Harrison, Duc, Trombe, Lord, Amato, and
  Baines}}]{mendels07}
\bibinfo{author}{\bibfnamefont{P.}~\bibnamefont{Mendels}},
  \bibinfo{author}{\bibfnamefont{F.}~\bibnamefont{Bert}},
  \bibinfo{author}{\bibfnamefont{M.}~\bibnamefont{de~Vries}},
  \bibinfo{author}{\bibfnamefont{A.}~\bibnamefont{Olariu}},
  \bibinfo{author}{\bibfnamefont{A.}~\bibnamefont{Harrison}},
  \bibinfo{author}{\bibfnamefont{F.}~\bibnamefont{Duc}},
  \bibinfo{author}{\bibfnamefont{J.}~\bibnamefont{Trombe}},
  \bibinfo{author}{\bibfnamefont{J.}~\bibnamefont{Lord}},
  \bibinfo{author}{\bibfnamefont{A.}~\bibnamefont{Amato}}, \bibnamefont{and}
  \bibinfo{author}{\bibfnamefont{C.}~\bibnamefont{Baines}},
  \bibinfo{journal}{Phys. Rev. Lett.} \textbf{\bibinfo{volume}{98}},
  \bibinfo{pages}{077204} (\bibinfo{year}{2007}).

\bibitem[{\citenamefont{Ofer et~al.}()\citenamefont{Ofer, Keren, Nytko, Shores,
  Bartlett, Nocera, Baines, and Amato}}]{ofer06}
\bibinfo{author}{\bibfnamefont{O.}~\bibnamefont{Ofer}},
  \bibinfo{author}{\bibfnamefont{A.}~\bibnamefont{Keren}},
  \bibinfo{author}{\bibfnamefont{E.~A.} \bibnamefont{Nytko}},
  \bibinfo{author}{\bibfnamefont{M.~P.} \bibnamefont{Shores}},
  \bibinfo{author}{\bibfnamefont{B.~M.} \bibnamefont{Bartlett}},
  \bibinfo{author}{\bibfnamefont{D.~G.} \bibnamefont{Nocera}},
  \bibinfo{author}{\bibfnamefont{C.}~\bibnamefont{Baines}}, \bibnamefont{and}
  \bibinfo{author}{\bibfnamefont{A.}~\bibnamefont{Amato}},
  \bibinfo{note}{arXiv:cond-mat/0610540 (2006)}.

\bibitem[{\citenamefont{Imai et~al.}(2008)\citenamefont{Imai, Nytko, Bartlett,
  Shores, and Nocera}}]{imai08}
\bibinfo{author}{\bibfnamefont{T.}~\bibnamefont{Imai}},
  \bibinfo{author}{\bibfnamefont{E.~A.} \bibnamefont{Nytko}},
  \bibinfo{author}{\bibfnamefont{B.}~\bibnamefont{Bartlett}},
  \bibinfo{author}{\bibfnamefont{M.}~\bibnamefont{Shores}}, \bibnamefont{and}
  \bibinfo{author}{\bibfnamefont{D.~G.} \bibnamefont{Nocera}},
  \bibinfo{journal}{Phys. Rev. Lett.} \textbf{\bibinfo{volume}{100}},
  \bibinfo{pages}{077203} (\bibinfo{year}{2008}).

\bibitem[{\citenamefont{Lee et~al.}(2007)\citenamefont{Lee, Kikuchi, Qiu, Lake,
  Huang, Habicht, and Kiefer}}]{shlee07}
\bibinfo{author}{\bibfnamefont{S.-H.} \bibnamefont{Lee}},
  \bibinfo{author}{\bibfnamefont{H.}~\bibnamefont{Kikuchi}},
  \bibinfo{author}{\bibfnamefont{Y.}~\bibnamefont{Qiu}},
  \bibinfo{author}{\bibfnamefont{B.}~\bibnamefont{Lake}},
  \bibinfo{author}{\bibfnamefont{Q.}~\bibnamefont{Huang}},
  \bibinfo{author}{\bibfnamefont{K.}~\bibnamefont{Habicht}}, \bibnamefont{and}
  \bibinfo{author}{\bibfnamefont{K.}~\bibnamefont{Kiefer}},
  \bibinfo{journal}{Nature Materials} \textbf{\bibinfo{volume}{6}},
  \bibinfo{pages}{853} (\bibinfo{year}{2007}).

\bibitem[{\citenamefont{de~Vries et~al.}(2008)\citenamefont{de~Vries, Kamenev,
  Kockelmann, Sanchez-Benitez, and Harrison}}]{devries07}
\bibinfo{author}{\bibfnamefont{M.~A.} \bibnamefont{de~Vries}},
  \bibinfo{author}{\bibfnamefont{K.~V.} \bibnamefont{Kamenev}},
  \bibinfo{author}{\bibfnamefont{W.~A.} \bibnamefont{Kockelmann}},
  \bibinfo{author}{\bibfnamefont{J.}~\bibnamefont{Sanchez-Benitez}},
  \bibnamefont{and} \bibinfo{author}{\bibfnamefont{A.}~\bibnamefont{Harrison}},
  \bibinfo{journal}{Phys. Rev. Lett.} \textbf{\bibinfo{volume}{100}},
  \bibinfo{pages}{157205} (\bibinfo{year}{2008}).

\bibitem[{\citenamefont{Bert et~al.}(2007)\citenamefont{Bert, Nakamae, Ladieu,
  L'Hote, Bonville, Duc, Trombe, and Mendels}}]{bert07}
\bibinfo{author}{\bibfnamefont{F.}~\bibnamefont{Bert}},
  \bibinfo{author}{\bibfnamefont{S.}~\bibnamefont{Nakamae}},
  \bibinfo{author}{\bibfnamefont{F.}~\bibnamefont{Ladieu}},
  \bibinfo{author}{\bibfnamefont{D.}~\bibnamefont{L'Hote}},
  \bibinfo{author}{\bibfnamefont{P.}~\bibnamefont{Bonville}},
  \bibinfo{author}{\bibfnamefont{F.}~\bibnamefont{Duc}},
  \bibinfo{author}{\bibfnamefont{J.-C.} \bibnamefont{Trombe}},
  \bibnamefont{and} \bibinfo{author}{\bibfnamefont{P.}~\bibnamefont{Mendels}},
  \bibinfo{journal}{Phys. Rev. B} \textbf{\bibinfo{volume}{76}},
  \bibinfo{pages}{132411} (\bibinfo{year}{2007}).

\bibitem[{\citenamefont{Olariu et~al.}(2008)\citenamefont{Olariu, Mendels,
  Bert, Duc, Trombe, de~Vries, and Harrison}}]{olariu08}
\bibinfo{author}{\bibfnamefont{A.}~\bibnamefont{Olariu}},
  \bibinfo{author}{\bibfnamefont{P.}~\bibnamefont{Mendels}},
  \bibinfo{author}{\bibfnamefont{F.}~\bibnamefont{Bert}},
  \bibinfo{author}{\bibfnamefont{F.}~\bibnamefont{Duc}},
  \bibinfo{author}{\bibfnamefont{J.~C.} \bibnamefont{Trombe}},
  \bibinfo{author}{\bibfnamefont{M.~A.} \bibnamefont{de~Vries}},
  \bibnamefont{and} \bibinfo{author}{\bibfnamefont{A.}~\bibnamefont{Harrison}},
  \bibinfo{journal}{Phys. Rev. Lett.} \textbf{\bibinfo{volume}{100}},
  \bibinfo{pages}{087202} (\bibinfo{year}{2008}).

\bibitem[{\citenamefont{Anderson}(1973)}]{anderson73}
\bibinfo{author}{\bibfnamefont{P.~W.} \bibnamefont{Anderson}},
  \bibinfo{journal}{Mat. Res. Bull.} \textbf{\bibinfo{volume}{8}},
  \bibinfo{pages}{153} (\bibinfo{year}{1973}).

\bibitem[{\citenamefont{Anderson}(1987)}]{anderson87}
\bibinfo{author}{\bibfnamefont{P.~W.} \bibnamefont{Anderson}},
  \bibinfo{journal}{Science} \textbf{\bibinfo{volume}{235}},
  \bibinfo{pages}{1196} (\bibinfo{year}{1987}).

\bibitem[{\citenamefont{Marston and Zeng}(1991)}]{marston91}
\bibinfo{author}{\bibfnamefont{J.~B.} \bibnamefont{Marston}} \bibnamefont{and}
  \bibinfo{author}{\bibfnamefont{C.}~\bibnamefont{Zeng}}, \bibinfo{journal}{J.
  Appl. Phys.} \textbf{\bibinfo{volume}{69}}, \bibinfo{pages}{5962}
  (\bibinfo{year}{1991}).

\bibitem[{\citenamefont{Sachdev}(1992)}]{sachdev92}
\bibinfo{author}{\bibfnamefont{S.}~\bibnamefont{Sachdev}},
  \bibinfo{journal}{Phys. Rev. B} \textbf{\bibinfo{volume}{45}},
  \bibinfo{pages}{12377} (\bibinfo{year}{1992}).

\bibitem[{\citenamefont{Hastings}(2000)}]{hastings00}
\bibinfo{author}{\bibfnamefont{M.~B.} \bibnamefont{Hastings}},
  \bibinfo{journal}{Phys. Rev. B} \textbf{\bibinfo{volume}{63}},
  \bibinfo{pages}{014413} (\bibinfo{year}{2000}).

\bibitem[{\citenamefont{Nikolic and Senthil}(2003)}]{nikolic03}
\bibinfo{author}{\bibfnamefont{P.}~\bibnamefont{Nikolic}} \bibnamefont{and}
  \bibinfo{author}{\bibfnamefont{T.}~\bibnamefont{Senthil}},
  \bibinfo{journal}{Phys. Rev. B} \textbf{\bibinfo{volume}{68}},
  \bibinfo{pages}{214415} (\bibinfo{year}{2003}).

\bibitem[{\citenamefont{Leung and Elser}(1993)}]{leung93}
\bibinfo{author}{\bibfnamefont{P.~W.} \bibnamefont{Leung}} \bibnamefont{and}
  \bibinfo{author}{\bibfnamefont{V.}~\bibnamefont{Elser}},
  \bibinfo{journal}{Phys. Rev. B} \textbf{\bibinfo{volume}{47}},
  \bibinfo{pages}{5459} (\bibinfo{year}{1993}).

\bibitem[{\citenamefont{Elstner and Young}(1994)}]{elstner94}
\bibinfo{author}{\bibfnamefont{N.}~\bibnamefont{Elstner}} \bibnamefont{and}
  \bibinfo{author}{\bibfnamefont{A.~P.} \bibnamefont{Young}},
  \bibinfo{journal}{Phys. Rev. B} \textbf{\bibinfo{volume}{50}},
  \bibinfo{pages}{6871} (\bibinfo{year}{1994}).

\bibitem[{\citenamefont{Zeng and Elser}(1995)}]{zeng95}
\bibinfo{author}{\bibfnamefont{C.}~\bibnamefont{Zeng}} \bibnamefont{and}
  \bibinfo{author}{\bibfnamefont{V.}~\bibnamefont{Elser}},
  \bibinfo{journal}{Phys. Rev. B} \textbf{\bibinfo{volume}{51}},
  \bibinfo{pages}{8318} (\bibinfo{year}{1995}).

\bibitem[{\citenamefont{Lecheminant et~al.}(1997)\citenamefont{Lecheminant,
  Bernu, Lhuillier, Pierre, and Sindzingre}}]{lecheminant97}
\bibinfo{author}{\bibfnamefont{P.}~\bibnamefont{Lecheminant}},
  \bibinfo{author}{\bibfnamefont{B.}~\bibnamefont{Bernu}},
  \bibinfo{author}{\bibfnamefont{C.}~\bibnamefont{Lhuillier}},
  \bibinfo{author}{\bibfnamefont{L.}~\bibnamefont{Pierre}}, \bibnamefont{and}
  \bibinfo{author}{\bibfnamefont{P.}~\bibnamefont{Sindzingre}},
  \bibinfo{journal}{Phys. Rev. B} \textbf{\bibinfo{volume}{56}},
  \bibinfo{pages}{2521} (\bibinfo{year}{1997}).

\bibitem[{\citenamefont{Waldtmann et~al.}(1998)\citenamefont{Waldtmann, Everts,
  Bernu, Lhuillier, Sindzingre, and Pierre}}]{waldtmann98}
\bibinfo{author}{\bibfnamefont{C.}~\bibnamefont{Waldtmann}},
  \bibinfo{author}{\bibfnamefont{H.-U.} \bibnamefont{Everts}},
  \bibinfo{author}{\bibfnamefont{B.}~\bibnamefont{Bernu}},
  \bibinfo{author}{\bibfnamefont{C.}~\bibnamefont{Lhuillier}},
  \bibinfo{author}{\bibfnamefont{P.}~\bibnamefont{Sindzingre}},
  \bibnamefont{and} \bibinfo{author}{\bibfnamefont{L.}~\bibnamefont{Pierre}},
  \bibinfo{journal}{Eur. Phys. J. B} \textbf{\bibinfo{volume}{2}},
  \bibinfo{pages}{501} (\bibinfo{year}{1998}).

\bibitem[{\citenamefont{Sindzingre et~al.}(2000)\citenamefont{Sindzingre,
  Misguich, Lhuillier, Bernu, Pierre, Waldtmann, and Everts}}]{sindzingre00}
\bibinfo{author}{\bibfnamefont{P.}~\bibnamefont{Sindzingre}},
  \bibinfo{author}{\bibfnamefont{G.}~\bibnamefont{Misguich}},
  \bibinfo{author}{\bibfnamefont{C.}~\bibnamefont{Lhuillier}},
  \bibinfo{author}{\bibfnamefont{B.}~\bibnamefont{Bernu}},
  \bibinfo{author}{\bibfnamefont{L.}~\bibnamefont{Pierre}},
  \bibinfo{author}{\bibfnamefont{C.}~\bibnamefont{Waldtmann}},
  \bibnamefont{and} \bibinfo{author}{\bibfnamefont{H.-U.}
  \bibnamefont{Everts}}, \bibinfo{journal}{Phys. Rev. Lett.}
  \textbf{\bibinfo{volume}{84}}, \bibinfo{pages}{2953} (\bibinfo{year}{2000}).

\bibitem[{\citenamefont{Misguich and Sindzingre}(2007)}]{misguich07}
\bibinfo{author}{\bibfnamefont{G.}~\bibnamefont{Misguich}} \bibnamefont{and}
  \bibinfo{author}{\bibfnamefont{P.}~\bibnamefont{Sindzingre}},
  \bibinfo{journal}{Eur. Phys. J. B} \textbf{\bibinfo{volume}{59}},
  \bibinfo{pages}{305} (\bibinfo{year}{2007}).

\bibitem[{\citenamefont{Singh and Huse}(2007)}]{singh07}
\bibinfo{author}{\bibfnamefont{R.~R.~P.} \bibnamefont{Singh}} \bibnamefont{and}
  \bibinfo{author}{\bibfnamefont{D.~A.} \bibnamefont{Huse}},
  \bibinfo{journal}{Phys. Rev. B} \textbf{\bibinfo{volume}{76}},
  \bibinfo{pages}{180407} (\bibinfo{year}{2007}).

\bibitem[{\citenamefont{Singh and Huse}(2008)}]{singh08}
\bibinfo{author}{\bibfnamefont{R.~R.~P.} \bibnamefont{Singh}} \bibnamefont{and}
  \bibinfo{author}{\bibfnamefont{D.~A.} \bibnamefont{Huse}},
  \bibinfo{journal}{Phys. Rev. B} \textbf{\bibinfo{volume}{77}},
  \bibinfo{pages}{144415} (\bibinfo{year}{2008}).

\bibitem[{\citenamefont{Yang et~al.}()\citenamefont{Yang, Kim, Yu, and
  Park}}]{yang08}
\bibinfo{author}{\bibfnamefont{B.-J.} \bibnamefont{Yang}},
  \bibinfo{author}{\bibfnamefont{Y.~B.} \bibnamefont{Kim}},
  \bibinfo{author}{\bibfnamefont{J.}~\bibnamefont{Yu}}, \bibnamefont{and}
  \bibinfo{author}{\bibfnamefont{K.}~\bibnamefont{Park}},
  \bibinfo{note}{arXiv:0802.4343 [cond-mat.str-el] (2008)}.

\bibitem[{\citenamefont{Misguich and Lhuillier}(2005)}]{misguich05}
\bibinfo{author}{\bibfnamefont{G.}~\bibnamefont{Misguich}} \bibnamefont{and}
  \bibinfo{author}{\bibfnamefont{C.}~\bibnamefont{Lhuillier}}, in
  \emph{\bibinfo{booktitle}{Frustrated spin systems}}, edited by
  \bibinfo{editor}{\bibfnamefont{H.~T.} \bibnamefont{Diep}}
  (\bibinfo{publisher}{World Scientific}, \bibinfo{year}{2005}), p.
  \bibinfo{pages}{229}.

\bibitem[{\citenamefont{Ran et~al.}(2007)\citenamefont{Ran, Hermele, Lee, and
  Wen}}]{ran07}
\bibinfo{author}{\bibfnamefont{Y.}~\bibnamefont{Ran}},
  \bibinfo{author}{\bibfnamefont{M.}~\bibnamefont{Hermele}},
  \bibinfo{author}{\bibfnamefont{P.~A.} \bibnamefont{Lee}}, \bibnamefont{and}
  \bibinfo{author}{\bibfnamefont{X.-G.} \bibnamefont{Wen}},
  \bibinfo{journal}{Phys. Rev. Lett.} \textbf{\bibinfo{volume}{98}},
  \bibinfo{pages}{117205} (\bibinfo{year}{2007}).

\bibitem[{\citenamefont{Rantner and Wen}(2001)}]{rantner01}
\bibinfo{author}{\bibfnamefont{W.}~\bibnamefont{Rantner}} \bibnamefont{and}
  \bibinfo{author}{\bibfnamefont{X.-G.} \bibnamefont{Wen}},
  \bibinfo{journal}{Phys. Rev. Lett.} \textbf{\bibinfo{volume}{86}},
  \bibinfo{pages}{3871} (\bibinfo{year}{2001}).

\bibitem[{\citenamefont{Rantner and Wen}(2002)}]{rantner02}
\bibinfo{author}{\bibfnamefont{W.}~\bibnamefont{Rantner}} \bibnamefont{and}
  \bibinfo{author}{\bibfnamefont{X.-G.} \bibnamefont{Wen}},
  \bibinfo{journal}{Phys. Rev. B} \textbf{\bibinfo{volume}{66}},
  \bibinfo{pages}{144501} (\bibinfo{year}{2002}).

\bibitem[{\citenamefont{Ran et~al.}()\citenamefont{Ran, Ko, Lee, and
  Wen}}]{ran07b}
\bibinfo{author}{\bibfnamefont{Y.}~\bibnamefont{Ran}},
  \bibinfo{author}{\bibfnamefont{W.-H.} \bibnamefont{Ko}},
  \bibinfo{author}{\bibfnamefont{P.~A.} \bibnamefont{Lee}}, \bibnamefont{and}
  \bibinfo{author}{\bibfnamefont{X.-G.} \bibnamefont{Wen}},
  \bibinfo{note}{arXiv:0710.4574 [cond-mat.str-el] (2007)}.

\bibitem[{\citenamefont{Gregor and Motrunich}(2008)}]{gregor08}
\bibinfo{author}{\bibfnamefont{K.}~\bibnamefont{Gregor}} \bibnamefont{and}
  \bibinfo{author}{\bibfnamefont{O.~I.} \bibnamefont{Motrunich}},
  \bibinfo{journal}{Phys. Rev. B} \textbf{\bibinfo{volume}{77}},
  \bibinfo{pages}{184423} (\bibinfo{year}{2008}).

\bibitem[{\citenamefont{Ma and Marston}()}]{ma08}
\bibinfo{author}{\bibfnamefont{O.}~\bibnamefont{Ma}} \bibnamefont{and}
  \bibinfo{author}{\bibfnamefont{J.~B.} \bibnamefont{Marston}},
  \bibinfo{note}{arXiv:0801.2138 [cond-mat.str-el] (2008)}.

\bibitem[{\citenamefont{Ryu et~al.}(2007)\citenamefont{Ryu, Motrunich, Alicea,
  and Fisher}}]{ryu07}
\bibinfo{author}{\bibfnamefont{S.}~\bibnamefont{Ryu}},
  \bibinfo{author}{\bibfnamefont{O.~I.} \bibnamefont{Motrunich}},
  \bibinfo{author}{\bibfnamefont{J.}~\bibnamefont{Alicea}}, \bibnamefont{and}
  \bibinfo{author}{\bibfnamefont{M.~P.~A.} \bibnamefont{Fisher}},
  \bibinfo{journal}{Phys. Rev. B} \textbf{\bibinfo{volume}{75}},
  \bibinfo{pages}{184406} (\bibinfo{year}{2007}).

\bibitem[{\citenamefont{Borokhov et~al.}(2002)\citenamefont{Borokhov, Kapustin,
  and Wu}}]{borokhov02}
\bibinfo{author}{\bibfnamefont{V.}~\bibnamefont{Borokhov}},
  \bibinfo{author}{\bibfnamefont{A.}~\bibnamefont{Kapustin}}, \bibnamefont{and}
  \bibinfo{author}{\bibfnamefont{X.}~\bibnamefont{Wu}}, \bibinfo{journal}{J.
  High Energy Phys.} \textbf{\bibinfo{volume}{2002}}, \bibinfo{pages}{049}
  (\bibinfo{year}{2002}).

\bibitem[{\citenamefont{Vafek et~al.}(2002)\citenamefont{Vafek, Tesanovic, and
  Franz}}]{vafek02}
\bibinfo{author}{\bibfnamefont{O.}~\bibnamefont{Vafek}},
  \bibinfo{author}{\bibfnamefont{Z.}~\bibnamefont{Tesanovic}},
  \bibnamefont{and} \bibinfo{author}{\bibfnamefont{M.}~\bibnamefont{Franz}},
  \bibinfo{journal}{Phys. Rev. Lett.} \textbf{\bibinfo{volume}{89}},
  \bibinfo{pages}{157003} (\bibinfo{year}{2002}).

\bibitem[{\citenamefont{Hermele et~al.}(2004)\citenamefont{Hermele, Senthil,
  Fisher, Lee, Nagaosa, and Wen}}]{hermele04}
\bibinfo{author}{\bibfnamefont{M.}~\bibnamefont{Hermele}},
  \bibinfo{author}{\bibfnamefont{T.}~\bibnamefont{Senthil}},
  \bibinfo{author}{\bibfnamefont{M.~P.~A.} \bibnamefont{Fisher}},
  \bibinfo{author}{\bibfnamefont{P.~A.} \bibnamefont{Lee}},
  \bibinfo{author}{\bibfnamefont{N.}~\bibnamefont{Nagaosa}}, \bibnamefont{and}
  \bibinfo{author}{\bibfnamefont{X.-G.} \bibnamefont{Wen}},
  \bibinfo{journal}{Phys. Rev. B} \textbf{\bibinfo{volume}{70}},
  \bibinfo{pages}{214437} (\bibinfo{year}{2004}).

\bibitem[{\citenamefont{Hermele et~al.}(2005)\citenamefont{Hermele, Senthil,
  and Fisher}}]{hermele05}
\bibinfo{author}{\bibfnamefont{M.}~\bibnamefont{Hermele}},
  \bibinfo{author}{\bibfnamefont{T.}~\bibnamefont{Senthil}}, \bibnamefont{and}
  \bibinfo{author}{\bibfnamefont{M.~P.~A.} \bibnamefont{Fisher}},
  \bibinfo{journal}{Phys. Rev. B} \textbf{\bibinfo{volume}{72}},
  \bibinfo{pages}{104404} (\bibinfo{year}{2005}).

\bibitem[{\citenamefont{Alicea et~al.}(2005{\natexlab{a}})\citenamefont{Alicea,
  Motrunich, Hermele, and Fisher}}]{alicea05a}
\bibinfo{author}{\bibfnamefont{J.}~\bibnamefont{Alicea}},
  \bibinfo{author}{\bibfnamefont{O.~I.} \bibnamefont{Motrunich}},
  \bibinfo{author}{\bibfnamefont{M.}~\bibnamefont{Hermele}}, \bibnamefont{and}
  \bibinfo{author}{\bibfnamefont{M.~P.~A.} \bibnamefont{Fisher}},
  \bibinfo{journal}{Phys. Rev. B} \textbf{\bibinfo{volume}{72}},
  \bibinfo{pages}{064407} (\bibinfo{year}{2005}{\natexlab{a}}).

\bibitem[{\citenamefont{Alicea et~al.}(2005{\natexlab{b}})\citenamefont{Alicea,
  Motrunich, and Fisher}}]{alicea05b}
\bibinfo{author}{\bibfnamefont{J.}~\bibnamefont{Alicea}},
  \bibinfo{author}{\bibfnamefont{O.~I.} \bibnamefont{Motrunich}},
  \bibnamefont{and} \bibinfo{author}{\bibfnamefont{M.~P.~A.}
  \bibnamefont{Fisher}}, \bibinfo{journal}{Phys. Rev. Lett.}
  \textbf{\bibinfo{volume}{95}}, \bibinfo{pages}{247203}
  (\bibinfo{year}{2005}{\natexlab{b}}).

\bibitem[{\citenamefont{Alicea et~al.}(2006)\citenamefont{Alicea, Motrunich,
  and Fisher}}]{alicea06}
\bibinfo{author}{\bibfnamefont{J.}~\bibnamefont{Alicea}},
  \bibinfo{author}{\bibfnamefont{O.~I.} \bibnamefont{Motrunich}},
  \bibnamefont{and} \bibinfo{author}{\bibfnamefont{M.~P.~A.}
  \bibnamefont{Fisher}}, \bibinfo{journal}{Phys. Rev. B}
  \textbf{\bibinfo{volume}{73}}, \bibinfo{pages}{174430}
  (\bibinfo{year}{2006}).

\bibitem[{\citenamefont{Rigol and Singh}(2007{\natexlab{a}})}]{rigol07a}
\bibinfo{author}{\bibfnamefont{M.}~\bibnamefont{Rigol}} \bibnamefont{and}
  \bibinfo{author}{\bibfnamefont{R.~R.~P.} \bibnamefont{Singh}},
  \bibinfo{journal}{Phys. Rev. Lett.} \textbf{\bibinfo{volume}{98}},
  \bibinfo{pages}{207204} (\bibinfo{year}{2007}{\natexlab{a}}).

\bibitem[{\citenamefont{Rigol and Singh}(2007{\natexlab{b}})}]{rigol07b}
\bibinfo{author}{\bibfnamefont{M.}~\bibnamefont{Rigol}} \bibnamefont{and}
  \bibinfo{author}{\bibfnamefont{R.~R.~P.} \bibnamefont{Singh}},
  \bibinfo{journal}{Phys. Rev. B} \textbf{\bibinfo{volume}{76}},
  \bibinfo{pages}{184403} (\bibinfo{year}{2007}{\natexlab{b}}).

\bibitem[{\citenamefont{Appelquist et~al.}(1986)\citenamefont{Appelquist,
  Bowick, Karabali, and Wijewardhana}}]{appelquist86}
\bibinfo{author}{\bibfnamefont{T.~W.} \bibnamefont{Appelquist}},
  \bibinfo{author}{\bibfnamefont{M.}~\bibnamefont{Bowick}},
  \bibinfo{author}{\bibfnamefont{D.}~\bibnamefont{Karabali}}, \bibnamefont{and}
  \bibinfo{author}{\bibfnamefont{L.~C.~R.} \bibnamefont{Wijewardhana}},
  \bibinfo{journal}{Phys. Rev. D} \textbf{\bibinfo{volume}{33}},
  \bibinfo{pages}{3704} (\bibinfo{year}{1986}).

\bibitem[{\citenamefont{Franz et~al.}(2002)\citenamefont{Franz, Tesanovic, and
  Vafek}}]{franz02}
\bibinfo{author}{\bibfnamefont{M.}~\bibnamefont{Franz}},
  \bibinfo{author}{\bibfnamefont{Z.}~\bibnamefont{Tesanovic}},
  \bibnamefont{and} \bibinfo{author}{\bibfnamefont{O.}~\bibnamefont{Vafek}},
  \bibinfo{journal}{Phys. Rev. B} \textbf{\bibinfo{volume}{66}},
  \bibinfo{pages}{054535} (\bibinfo{year}{2002}).

\bibitem[{\citenamefont{Wen}(2002)}]{wen02}
\bibinfo{author}{\bibfnamefont{X.-G.} \bibnamefont{Wen}},
  \bibinfo{journal}{Phys. Rev. B} \textbf{\bibinfo{volume}{65}},
  \bibinfo{pages}{165113} (\bibinfo{year}{2002}).

\bibitem[{\citenamefont{Chubukov}(1992)}]{chubukov92}
\bibinfo{author}{\bibfnamefont{A.}~\bibnamefont{Chubukov}},
  \bibinfo{journal}{Phys. Rev. Lett.} \textbf{\bibinfo{volume}{69}},
  \bibinfo{pages}{832} (\bibinfo{year}{1992}).

\bibitem[{\citenamefont{Chandra et~al.}(1993)\citenamefont{Chandra, Coleman,
  and Ritchey}}]{chandra93}
\bibinfo{author}{\bibfnamefont{P.}~\bibnamefont{Chandra}},
  \bibinfo{author}{\bibfnamefont{P.}~\bibnamefont{Coleman}}, \bibnamefont{and}
  \bibinfo{author}{\bibfnamefont{I.}~\bibnamefont{Ritchey}},
  \bibinfo{journal}{J. de Physique I} \textbf{\bibinfo{volume}{3}},
  \bibinfo{pages}{591} (\bibinfo{year}{1993}).

\bibitem[{\citenamefont{Henley and Chan}(1995)}]{henley95}
\bibinfo{author}{\bibfnamefont{C.~L.} \bibnamefont{Henley}} \bibnamefont{and}
  \bibinfo{author}{\bibfnamefont{E.~P.} \bibnamefont{Chan}},
  \bibinfo{journal}{J. Magn. Magn. Mater.} \textbf{\bibinfo{volume}{140-144}},
  \bibinfo{pages}{1693} (\bibinfo{year}{1995}).

\bibitem[{\citenamefont{Cross and Fisher}(1979)}]{cross79}
\bibinfo{author}{\bibfnamefont{M.~C.} \bibnamefont{Cross}} \bibnamefont{and}
  \bibinfo{author}{\bibfnamefont{D.~S.} \bibnamefont{Fisher}},
  \bibinfo{journal}{Phys. Rev. B} \textbf{\bibinfo{volume}{19}},
  \bibinfo{pages}{402} (\bibinfo{year}{1979}).

\bibitem[{\citenamefont{Abel et~al.}(2007)\citenamefont{Abel, Matan, Chou,
  Isaacs, Moncton, Sinn, Alatas, and Lee}}]{abel07}
\bibinfo{author}{\bibfnamefont{E.~T.} \bibnamefont{Abel}},
  \bibinfo{author}{\bibfnamefont{K.}~\bibnamefont{Matan}},
  \bibinfo{author}{\bibfnamefont{F.~C.} \bibnamefont{Chou}},
  \bibinfo{author}{\bibfnamefont{E.~D.} \bibnamefont{Isaacs}},
  \bibinfo{author}{\bibfnamefont{D.~E.} \bibnamefont{Moncton}},
  \bibinfo{author}{\bibfnamefont{H.}~\bibnamefont{Sinn}},
  \bibinfo{author}{\bibfnamefont{A.}~\bibnamefont{Alatas}}, \bibnamefont{and}
  \bibinfo{author}{\bibfnamefont{Y.~S.} \bibnamefont{Lee}},
  \bibinfo{journal}{Phys. Rev. B} \textbf{\bibinfo{volume}{76}},
  \bibinfo{pages}{214304} (\bibinfo{year}{2007}).

\bibitem[{\citenamefont{Polyakov}(1977)}]{polyakov77}
\bibinfo{author}{\bibfnamefont{A.~M.} \bibnamefont{Polyakov}},
  \bibinfo{journal}{Nucl. Phys. B} \textbf{\bibinfo{volume}{120}},
  \bibinfo{pages}{429} (\bibinfo{year}{1977}).

\bibitem[{\citenamefont{Haldane}(1988)}]{haldane88}
\bibinfo{author}{\bibfnamefont{F.~D.~M.} \bibnamefont{Haldane}},
  \bibinfo{journal}{Phys. Rev. Lett.} \textbf{\bibinfo{volume}{61}},
  \bibinfo{pages}{2015} (\bibinfo{year}{1988}).

\bibitem[{\citenamefont{Gros}(1989)}]{gros89}
\bibinfo{author}{\bibfnamefont{C.}~\bibnamefont{Gros}}, \bibinfo{journal}{Ann.
  Phys.} \textbf{\bibinfo{volume}{189}}, \bibinfo{pages}{53}
  (\bibinfo{year}{1989}).

\bibitem[{\citenamefont{Kaul and Sachdev}(2008)}]{kaul08}
\bibinfo{author}{\bibfnamefont{R.~K.} \bibnamefont{Kaul}} \bibnamefont{and}
  \bibinfo{author}{\bibfnamefont{S.}~\bibnamefont{Sachdev}},
  \bibinfo{journal}{Phys. Rev. B} \textbf{\bibinfo{volume}{77}},
  \bibinfo{pages}{155105} (\bibinfo{year}{2008}).

\bibitem[{\citenamefont{Kolezhuk et~al.}(2006)\citenamefont{Kolezhuk, Sachdev,
  Biswas, and Chen}}]{kolezhuk06}
\bibinfo{author}{\bibfnamefont{A.}~\bibnamefont{Kolezhuk}},
  \bibinfo{author}{\bibfnamefont{S.}~\bibnamefont{Sachdev}},
  \bibinfo{author}{\bibfnamefont{R.~R.} \bibnamefont{Biswas}},
  \bibnamefont{and} \bibinfo{author}{\bibfnamefont{P.}~\bibnamefont{Chen}},
  \bibinfo{journal}{Phys. Rev. B} \textbf{\bibinfo{volume}{74}},
  \bibinfo{pages}{165114} (\bibinfo{year}{2006}).

\bibitem[{\citenamefont{Hamermesh}(1962)}]{hamermesh62}
\bibinfo{author}{\bibfnamefont{M.}~\bibnamefont{Hamermesh}},
  \emph{\bibinfo{title}{Group Theory and its Application to Physical
  Problems}}, Addison-Wesley Series in Physics
  (\bibinfo{publisher}{Addison-Wesley}, \bibinfo{year}{1962}).

\end{thebibliography}

\end{document}